\documentclass[lettersize,journal]{IEEEtran}
\usepackage{graphicx} 
\usepackage{xcolor}
\usepackage{algorithm}
\usepackage{algpseudocode}
\usepackage{bm}
\usepackage{subcaption}
\usepackage[T1]{fontenc}    
\usepackage{hyperref}       
\usepackage{url}            
\usepackage{booktabs}       
\usepackage{amsfonts}       
\usepackage{nicefrac}       
\usepackage{microtype}      
\usepackage{lipsum}		
\usepackage{graphicx}
\usepackage{doi}
\usepackage{amsmath}
\usepackage{verbatim}
\usepackage{multirow}
\usepackage[numbers]{natbib}
\newcommand{\nc}{N_{\text{c}}}
\def\BibTeX{{\rm B\kern-.05em{\sc i\kern-.025em b}\kern-.08em
    T\kern-.1667em\lower.7ex\hbox{E}\kern-.125emX}}
\usepackage{balance}
\begin{document}
\title{Accelerating Microswimmer Simulations via a Heterogeneous Pipelined Parallel-in-Time Framework}

\author{Ruixiang Huang, Weifan Liu
\thanks{This work is supported by National Natural Science Foundation of China (Grant number 12301547) and the Fundamental Research Funds for the Central Universities (Grant number BLX202245). (Corresponding author: Weifan Liu)}
\thanks{Ruixiang Huang is with Beijing Forestry University, Beijing, China, 100083 and University of Washington, Seattle, WA, USA 98195.}
\thanks{Weifan Liu is with Beijing Forestry University, Beijing, China, 100083}}

\markboth{IEEE Transactions on Parallel and Distributed Systems}%
{How to Use the IEEEtran \LaTeX \ Templates}

\maketitle
\begin{center}
\footnotesize
This work has been submitted to IEEE for possible publication. Copyright may be transferred without notice, after which this version may no longer be accessible.
\end{center}
\begin{abstract}
Simulating large-scale microswimmer dynamics in viscous fluid poses significant challenges due to the coupled high spatial and temporal complexity. Conventional high-performance computing (HPC) methods often address these two dimensions in isolation, leaving a critical gap for synergistic acceleration. This paper introduces a heterogeneous CPU--GPU computing framework specifically optimized for the long-time simulation of filamentous microswimmers in viscous fluid. We propose a two-level parallelization strategy: (1) high-intensity GPU kernels to resolve the quadratic spatial interactions given by the Method of Regularized Stokeslets (MRS), and (2) a distributed MPI-GPU pipelined Parareal architecture to exploit temporal concurrency. By mapping the asynchronous pipeline onto multiple GPU devices, our framework effectively overlaps coarse and fine propagators, overcoming the serial bottlenecks of traditional Parareal method. Furthermore, we employ a GPU-optimized numerical routine for computing the matrix square root arising in the numerical scheme of the filamentous microswimmer simulations. Theoretical analysis of the efficiency improvement of the pipelined Parareal is presented. Numerical experiments demonstrate that the proposed framework achieves order-of-magnitude speedups over CPU-only methods, providing a scalable pathway for simulating complex emergent behaviors in large-scale biology and physics systems.
\end{abstract}


\section{Introduction}
The locomotion of microorganisms, such as flagellated bacteria and sperm, is fundamentally governed by low-Reynolds-number dynamics, specifically Stokes flow. In computational models, these filamentous structures are typically modeled as thin elastic rods. The dynamic interplay between the rod's elastic properties and the fluid's viscous resistance forms the fundamental basis for studying biological propulsion and collective swimming behaviors \cite{Peskin2002}. The large-scale, long-term simulations of such systems are crucial for understanding the mechanisms of self-organization, emergence, and collective motion. However, the underlying fluid-structure interaction (FSI) problems remain extremely computationally expensive.

Initially, the Immersed Boundary (IB) method was developed to address such problems by coupling an Eulerian fluid grid with Lagrangian structural representations. 
However, this approach often necessitates substantial grid refinement to accurately resolve thin, filamentous structures, leading to a significant increase in computational overhead \cite{tornberg2004simulating}. As a mesh-free alternative, the Method of Regularized Stokeslets (MRS) was introduced, replacing singular Green's functions (Stokeslets) with smoothed, regularized kernels \cite{Cortez2001}. This regularization eliminates the singularity at the point forces, yielding a numerically stable and mathematically consistent framework for computing velocity fields induced by forces and moments distributed along slender bodies. Other notable techniques include boundary integral equation methods \cite{Pozrikidis1992} and the Rotne-Prager-Yamakawa (RPY) tensor \cite{Rotne1969, Yamakawa1970}, each offering distinct formulations for handling hydrodynamic interactions in slender structures. Despite the elegance and geometric versatility of these Lagrangian-based frameworks, they often rely on evaluating pairwise interactions. Consequently, the computational complexity scales quadratically with the number of discretization points, posing a primary bottleneck for large-scale simulations.

A secondary, yet equally daunting challenge lies in the temporal discretization. One approach to model the filamentous structures is to employ a version of the Kirchhoff rod \cite{carichino2019emergent, ho2019three, lim2012fluid}. The FSI problem of the Kirchhoff rod immersed in viscous fluid is inherently stiff. To maintain numerical stability, explicit numerical integrators, such as Runge-Kutta schemes, are often constrained by prohibitively small time steps. Consequently, simulating even a single stroke cycle may necessitate millions of iterations, rendering long-term evolutionary studies or large-scale collective dynamic simulations computationally intractable.

To break the sequential barrier of time integration, Parallel-in-Time (PinT) algorithms, which are a class of methods designed to parallelize the temporal dimension in the numerical solution of time-dependent differential equations, such as those arising in computational fluid dynamics, complex systems and multiscale physics simulations. Parareal, as one notable PinT method, has been proposed to distribute the temporal workload across multiple processing units \cite{lionsparareal}. By utilizing a coarse-grained predictor to provide an initial guess and a fine-grained propagator to achieve accuracy through iterative corrections, Parareal allows for the simultaneous calculation of multiple partitioned time intervals. In recent years, PinT methods have further evolved into multi-level time-parallel algorithms such as PFASST \cite{Emmett2012,freese2024psdc} and MGRIT \cite{falgout2014parallel,hahne2021atmgrit,margenberg2021fsi}, and have been systematically investigated in fields including fluid dynamics and multi-physics problems \cite{Ong2020PinTReview,BlumersEtAl2021,gander2026time,steinstraesser2024pint,liu2022parallel}.

To overcome the prohibitive computational overhead inherent in Lagrangian frameworks, such as the Method of Regularized Stokeslets (MRS), where pairwise interactions impose a significant computational load, GPU acceleration has emerged as a pivotal strategy. Leveraging the high-throughput data parallelism of GPUs, researchers have successfully shifted intensive tasks, including kernel evaluations and state updates, from the CPU to the GPU. Recent efforts in FSI have further demonstrated the efficacy of CPU–-GPU heterogeneous architectures, underscoring the potential of heterogeneous parallelism in tackling complex FSI challenges \cite{EGHBAL201757, Zeng2025, xue2024cpu, li2023incompressible}.

While a wide array of HPC algorithms have been developed for various fluid applications, specialized frameworks tailored for the large-scale, long-time simulation of filament dynamics remain remarkably scarce. Existing HPC efforts in this domain have predominantly focused on either spatial acceleration via GPU-based kernels \cite{gallagher2020passively} or temporal parallelism through CPU-based time-integration schemes \cite{liu2022parallel}. However, these decoupled approaches often fail to address the synergistic computational demands of biofluid simulations, which necessitate the simultaneous resolution of both high spatial and temporal complexities. Consequently, there remains a critical need for a specialized, heterogeneous CPU--GPU pipeline capable of concurrently managing spatial complexity and temporal concurrency to enable high-fidelity simulations of motile microswimmers.

In this paper, we address these multifaceted challenges by introducing a heterogeneous CPU--GPU computing framework specifically optimized for the filamentous dynamics characteristic of sperm and bacterial simulations in biofluid research. Our approach implements a two-level parallelization strategy. First, we leverage the massive throughput of GPUs to resolve the spatial complexity of MRS, utilizing custom-designed kernels for the high-throughput evaluation of linear and angular velocities. Concurrently, we deploy the Parareal algorithm across a CPU-based cluster to exploit temporal parallelism. By synergizing spatial acceleration with time-parallel integration, this framework mitigates the prohibitive computational cost of sequential time-integration for long-time biofluid simulations of collective phenomena and emergent behaviors.

Our primary contributions are summarized as follows:
\begin{itemize}
\item \textbf{A scalable multi-GPU pipelined Parareal framework:} We design a novel computing architecture that synergizes spatial and temporal parallelism by integrating a pipelined Parareal structure with a distributed MPI-GPU framework. By leveraging the asynchronous scheduling of the pipeline, our framework enables the concurrent utilization of multiple GPU devices, effectively overlapping coarse and fine solvers across clusters. This synergy allows the system to handle the $O(N^2)$ spatial complexity of Stokesian dynamics while simultaneously exploiting temporal concurrency, providing a scalable pathway that leverages abundant GPU resources to achieve higher utilization and lower GPU idle time in large-scale simulations.
    
\item \textbf{GPU-optimized numerical kernels for filament dynamics:} We develop high-intensity parallel kernels for the evaluation of linear and angular velocities, which represent the primary computational bottleneck due to the $O(N^2)$ pairwise interactions in Stokesian dynamics. Furthermore, we employ a numerical routine for calculating the rotation matrix square root---a critical yet sensitive operation for maintaining the orthonormal frame of Kirchhoff rods. This routine is explicitly optimized for the SIMT (Single Instruction, Multiple Threads) architecture, ensuring both rigorous numerical stability and high-performance execution of the long-time simulations of flagellar dynamics.
    
\item \textbf{Comprehensive performance analysis and validation:} We present a theoretical and empirical evaluation of the efficiency of the proposed framework. By analyzing GPU idle times and the dependence of speedup on the coarse-to-fine cost ratio and iteration counts, we validate the superiority of the pipelined multi-GPU approach over standard Parareal methods. Numerical experiments involving both single and multiple microswimmers demonstrate order-of-magnitude speedups compared to CPU-only methods.
\end{itemize}

The remainder of this paper is organized as follows. Section~II introduces the Kirchhoff rod model for flagellar dynamics, together with the governing fluid equations and numerical methods. Section~III describes the implementation of the proposed space--time parallel framework on a heterogeneous CPU--GPU architecture. Section~IV presents the experimental results and their analysis. Section~V concludes the paper with discussion.

\section{Preliminaries}

\subsection{Kirchhoff rod model}
To model the structure of the thin filamentous structure of sperm or bacteria flagella, we employ an unconstrained Kirchhoff rod model. The rod is represented by a centerline $\mathbf{X}(s, t)$ and an associated orthonormal material frame $\{ \mathbf{D}^1(s, t), \mathbf{D}^2(s, t), \mathbf{D}^3(s, t) \}$, where $s \in [0, L]$ is the Lagrangian parameter, initialized as the arclength in the reference state. The vectors $\mathbf{D}^1$ and $\mathbf{D}^2$ lie in the plane of the rod's cross-section, while $\mathbf{D}^3 = \partial \mathbf{X} / \partial s$ remains tangent to the centerline in the absence of shear. A schematic illlustration of the space curve representation of a single filament is shown in the lower left corner of Figure \ref{fig:diagram}. The balance of linear and angular momentum for a rod element in the viscosity-dominated regime (neglecting inertia) is given by:
\begin{equation}
\frac{\partial \mathbf{F}}{\partial s} + \mathbf{f} = 0, \qquad 
\frac{\partial \mathbf{N}}{\partial s} + \frac{\partial \mathbf{X}}{\partial s} \times \mathbf{F} + \mathbf{n} = 0,
\label{eq:momentum_balance}
\end{equation}
where $\mathbf{F}(s)$ and $\mathbf{N}(s)$ denote the internal contact force and moment, while $\mathbf{f}(s)$ and $\mathbf{n}(s)$ represent the external force and torque densities exerted by the fluid onto the rod. 

The constitutive relations for the internal force and moment are expressed in the material frame as $\mathbf{F} = \sum_{i=1}^3 F^i \mathbf{D}^i$ and $\mathbf{N} = \sum_{i=1}^3 N^i \mathbf{D}^i$:
\begin{equation}
N^i = a_i \left( \frac{\partial \mathbf{D}^j}{\partial s} \cdot \mathbf{D}^k - \Omega_i \right), 
F^i = b_i \left( \frac{\partial \mathbf{X}}{\partial s} \cdot \mathbf{D}^i - \delta_{3i} \right),
\label{eq:constitutive_N_F}
\end{equation}
where $(i, j, k)$ is a cyclic permutation of $(1, 2, 3)$. The coefficients $a_1, a_2$ are bending moduli, $a_3$ is the torsional rigidity, $b_1, b_2$ are shear moduli, and $b_3$ is the stretching stiffness. For an axisymmetric rod with a circular cross-section, we set $a_1 = a_2$ and $b_1 = b_2$. The vector $\boldsymbol{\Omega} = (\Omega_1, \Omega_2, \Omega_3)$ defines the intrinsic curvature and twist of the rod.

\subsection{Discretization of the Kirchhoff rod model}
To transform the continuous Kirchhoff model into a computationally tractable form, the rod centerline is spatially discretized. A rod of total length $L$ is divided into $M$ points with a uniform segment length $\Delta s = L/(M-1)$. The Lagrangian position of each point is defined as $s_k = (k-1) \Delta s$ for $k = 1, \dots, M$. Let $\mathbf{X}_k(t)$ denote the position vector of the $k$-th point, and $\{\mathbf{D}_k^1, \mathbf{D}_k^2, \mathbf{D}_k^3\}$ represent the orthonormal triad defining the local material frame at the point. Using a finite difference method, the discrete internal forces $\mathbf{F}_{k+1/2}$ and moments $\mathbf{N}_{k+1/2}$ are defined on the segments connecting nodes $k$ and $k+1$. The balance equations for the discrete force density $\mathbf{f}_k$ and torque density $\mathbf{n}_k$ at each node are given by:

\begin{align}
\mathbf{f}_k &= -\frac{\mathbf{F}_{k+\frac{1}{2}} - \mathbf{F}_{k-\frac{1}{2}}}{\Delta s},
\label{eq:discrete_f} \\
\mathbf{n}_k &= -\frac{\mathbf{N}_{k+\frac{1}{2}} - \mathbf{N}_{k-\frac{1}{2}}}{\Delta s} - \frac{1}{2} \Bigg( \frac{\mathbf{X}_{k+1} - \mathbf{X}_k}{\Delta s} \times \mathbf{F}_{k+\frac{1}{2}} \notag \\
&\qquad + \frac{\mathbf{X}_k - \mathbf{X}_{k-1}}{\Delta s} \times \mathbf{F}_{k-\frac{1}{2}} \Bigg),
\label{eq:discrete_n}
\end{align}

The internal force and moment components $F^j_{k+1/2}$ and $N^j_{k+1/2}$ are computed based on the discrete strains between adjacent nodes:
\begin{align}
F_{k+\frac{1}{2}}^{j} &= b_j \left( \frac{\mathbf{X}_{k+1} - \mathbf{X}_k}{\Delta s} \cdot \mathbf{D}_{k+\frac{1}{2}}^{j} - \delta_{3j} \right), \label{eq:F_comp_discrete} \\
N_{k+\frac{1}{2}}^{i} &= a_i \left( \frac{\mathbf{D}_{k+1}^j - \mathbf{D}_k^j}{\Delta s} \cdot \mathbf{D}_{k+\frac{1}{2}}^k - \Omega_i \right), \label{eq:N_comp_discrete}
\end{align}
where $(i, j, k)$ is a cyclic permutation of $(1, 2, 3)$ and $(\Omega_1, \Omega_2, \Omega_3)$ is the strain twist vector describing how the rod bends and twists. The mid-segment material frame $\mathbf{D}_{k+1/2}^{i}$ is determined based on the rotation between adjacent nodes. Specifically, we define the rotation matrix $\mathbf{A}_k$ that maps node $k$ to node $k+1$:
\begin{equation}
\mathbf{A}_k = \sum_{j=1}^3 \mathbf{D}_{k+1}^j (\mathbf{D}_k^j)^T, \quad \mathbf{D}_{k+\frac{1}{2}}^i = \mathbf{A}_k^{1/2} \mathbf{D}_k^i,
\label{eq:mid_triad}
\end{equation}
where $\mathbf{A}_k^{1/2}$ represents the rotation through half the angle between the two adjacent frames, ensuring a second-order accurate spatial discretization.

\subsection{Mathematical model of Stokes flow}

The locomotion of microorganisms occurs in a viscosity-dominated regime where the Reynolds number is nearly zero, rendering inertial effects negligible. The fluid dynamics is governed by the incompressible Stokes equations:
\begin{align}
    \mu \Delta \mathbf{u} - \nabla p + \mathbf{f}_{ext} &= 0, \label{eq:stokes_momentum} \\
    \nabla \cdot \mathbf{u} &= 0, \label{eq:stokes_continuity}
\end{align}
where $\mu$ is the dynamic viscosity, $\mathbf{u}$ is the velocity field, $p$ is the pressure, and $\mathbf{f}_{ext}$ represents the external force density exerted by the microswimmer on the fluid. 

To solve these equations numerically without encountering the singularities associated with point forces (Stokeslets), we employ MRS. In this framework, the singular Dirac delta distribution is replaced by a smooth regularization kernel $\phi_{\varepsilon}(\mathbf{r})$, known as a blob function. A common choice for the kernel, which we adopt here, is:
\begin{equation}
    \phi_{\varepsilon}(r) = \frac{15\varepsilon^4}{8\pi(r^2 + \varepsilon^2)^{7/2}},
\end{equation}
where $r = |\mathbf{x} - \mathbf{x}_0|$ and $\varepsilon$ is the regularization parameter representing the physical radius of the rod. 

The velocity $\mathbf{u}$ and angular velocity $\boldsymbol{\omega}$ at a position $\mathbf{x}$ induced by a force $\mathbf{f}_k$ and torque $\mathbf{n}_k$ concentrated at $\mathbf{X}_k$ are derived from the regularized Green's function $G_\varepsilon$ and the biharmonic function $B_\varepsilon$:
\begin{align}
    \mu \mathbf{u}(\mathbf{x}) &= \mathbf{f}_k H_1(r) + [(\mathbf{f}_k \cdot \mathbf{r})\mathbf{r}] H_2(r) + (\mathbf{n}_k \times \mathbf{r}) H_3(r), \label{eq:velocity_field} \\
    \mu \boldsymbol{\omega}(\mathbf{x}) &= (\mathbf{f}_k \times \mathbf{r}) H_3(r) + \mathbf{n}_k H_4(r) + [(\mathbf{n}_k \cdot \mathbf{r})\mathbf{r}] H_5(r), \label{eq:angular_velocity_field}
\end{align}
where the scalar functions $H_1$ through $H_5$ are analytical expressions derived from the choice of $\phi_\varepsilon$. The specific formulations of the $H_1$ and $H_2$ are given in \cite{olson2013modeling}.

By the principle of linear superposition, the total velocities at any node $i$ are obtained by a matrix-vector product of the form
\begin{equation}
    \begin{bmatrix} \mathbf{u}_i \\ \boldsymbol{\omega}_i \end{bmatrix} = \frac{1}{\mu} \sum_{j=1}^{N} \mathbb{M}_{ij} \begin{bmatrix} \mathbf{f}_j \\ \mathbf{n}_j \end{bmatrix},
    \label{eq:superposition}
\end{equation}
where $\mathbb{M}_{ij}$ is the mobility matrix. The $O(N^2)$ nature of this summation allows for straightforward parallelization, as the contribution of each node $j$ to node $i$ can be computed independently.

\subsection{Parareal Algorithm}\label{sec:parareal}
The Parareal algorithm is a parallel-in-time algorithm. The algorithm is designed to solve time-dependent problems by parallelizing the time domain of the solution. Consider an Initial Value Problem (IVP) of the following form:

\begin{equation}\label{eq:ode}
\frac{dx}{dt}=u(t,x)~~\text{for}~~t\in(0,T]~~\text{and}~~{x}(0)={x}_0.
\end{equation}

Consider a parallel computing environment with $\nc$ cores. In the Parareal algorithm, the time interval $[0,T]$ is uniformly split into $\nc$ sub-intervals, each of length $\Delta T = T/\nc$, and the time points are denoted as $T_n = n\Delta T$ for $n = 0, 1, 2, \dots, \nc$. The algorithm achieves parallelism through an iterative procedure that alternates between two types of computations: sequential calculation using a low-accuracy coarse integrator and concurrent calculation using a high-accuracy fine integrator. The fine solver typically employs a higher-order numerical scheme with a finer temporal resolution, thus providing greater accuracy than its coarse counterpart. We denote the coarse integrator as $\mathcal{G}(T_n, T_{n-1}, x_{n-1}^k)$ and the fine integrator as $\mathcal{F}(T_n, T_{n-1}, X_{n-1}^k)$, both of which are used to advance the solution of the initial value problem (IVP).

\begin{equation}\label{eq:oden}
\frac{dx}{dt} = u(t, x), \quad x(T_{n-1}) = x_{n-1}^k,
\end{equation}
for $t\in(T_{n-1}, T_n]$, $n=1,2,\cdots,\nc$. The procedure of a standard Parareal algorithm for solving Equation \eqref{eq:ode} is given below:
\begin{enumerate}
\item For $n=k,~k+1,~k+2,~\cdots,~\nc$, compute in serial 
\begin{equation}\label{eq:initial}
X^0_{n}=\mathcal{G}(T_{n},T_{n-1},X^0_{n-1}), \quad X_0^0=x_0.
\end{equation}
\item For $n=k,~k+1,~k+2,~\cdots,~\nc$, compute in parallel
\begin{equation}\label{eq:intermediate}
{X_n^k}^{\prime}=\mathcal{F}\left(T_{n},T_{n-1},X^{k-1}_{n-1}\right) .
\end{equation}
\item Let $X_n^k=X_n^{k-1}$ for $n=1,~2,~\cdots,~k-1$ and $X_k^k = {X_k^k}^{\prime}$. For $n=k+1,~k+2,~\cdots,~\nc$, correct ${X_n^k}^{\prime}$ by applying the coarse solver sequentially as follows:
\begin{equation}\label{eq:serial_correct}
X^{k}_{n}={X_n^k}^{\prime}+\mathcal{G}\left(T_{n},T_{n-1},X^{k}_{n-1}\right)-\mathcal{G}\left(T_{n},T_{n-1},X^{k-1}_{n-1}\right),
\end{equation}
\end{enumerate}
where $\mathcal{G}\left(T_{n},T_{n-1},X^{k-1}_{n-1}\right)$ in Eq. (\ref{eq:serial_correct}) has been computed in the previous iteration.

\section{GPU-accelerated heterogeneous implementation}
\label{sec:gpu_impl}
\begin{figure*}[t]
    \centering
    \includegraphics[width=.9\linewidth]{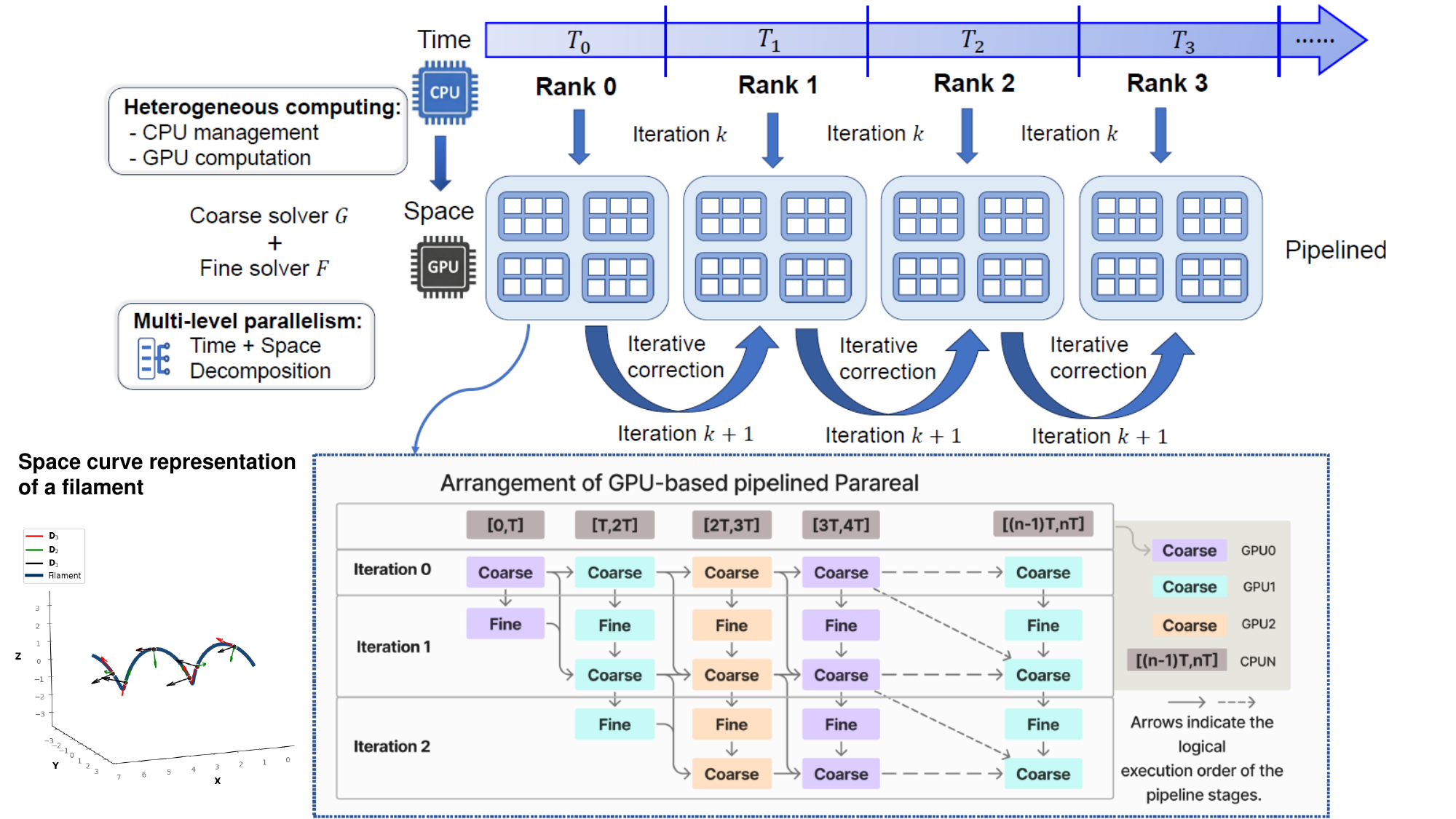}
    \caption{Schematic diagram of the proposed parallel computing pipeline}
    \label{fig:diagram}
\end{figure*}

\subsection{Pipelined parareal time parallelization}

In the classic Parareal method, the computational time domain is partitioned into $n$ subintervals, corresponding to $n$ parallel workers (e.g., CPU cores or processes). Although the fine solver can be executed fully in parallel over these subintervals, the coarse solver must still be advanced sequentially at the beginning of each iteration, which constitutes a clear performance bottleneck.
\begin{figure}[h]
    \centering
    \includegraphics[width=\linewidth]{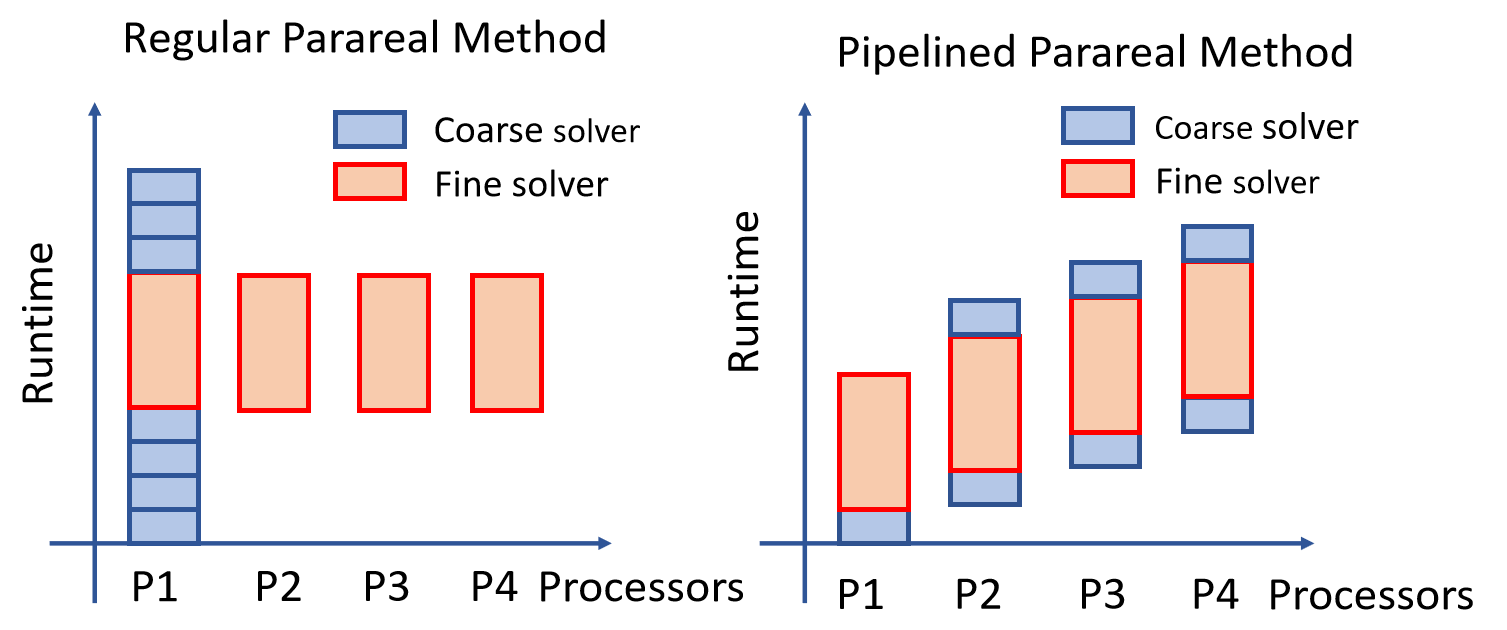}
    \caption{Schematic comparison of the computation flow of a regular Parareal method and a pipelined Parareal method. }
    \label{fig:parareal_compare}
\end{figure}
To alleviate this issue, we employ a pipelined scheduling strategy. Let the global time interval be decomposed into $N$ segments, where the $p$-th subinterval is assigned to the $p$-th worker. In the regular Parareal scheme, all workers must wait until the coarse solver completes the entire serial prediction from subinterval $0$ to $n-1$ before launching the fine solver. In contrast, under the pipelined strategy, worker $p$ can immediately start the fine solver as soon as the coarse solver finishes the prediction on the $p$th subinterval.

In other words, the coarse solver no longer needs to finish the entire serial propagation before the fine solver starts; instead, it streams intermediate results downstream as they become available. This allows each worker to begin fine solves as soon as it receives the required local initial condition, eliminating the inherent serial waiting. Such a strategy is especially beneficial on architectures with abundant CPU and GPU resources, as it significantly reduces GPU idle time and improves hardware utilization. Figure \ref{fig:parareal_compare} compares the computation flow of the regular and pipelined Parareal approaches. Figure \ref{fig:diagram} presents an overall schematic of the pipeline combining spatial and temporal parallelization on a CPU-GPU architecture, with the bottom panel illustrating the GPU arrangement strategy in the pipelined Parareal.

\subsubsection{Cost analysis of the pipelined Parareal}\label{ssec:cost}
We now quantitatively analyze the impact of different scheduling strategies on the overall runtime. When synchronization barriers exist or the wavefront has not fully developed, some GPUs remain idle.
For given computational workload and computing capability of GPUs, earlier participation of GPUs in computation generally leads to a shorter completion time. Therefore, we characterize the difference between the pipeline and regular scheduling strategies by analyzing GPU idle time. To describe the idleness, we define the total GPU wait time $W$, the sum of the individual waiting periods for each GPU, reflecting the extent of resource underutilization due to synchronization overheads or load imbalance. A lower value indicates better parallel efficiency and minimal resource stalling.
\begin{equation}
W = \sum_{i=1}^{m} (\text{GPU}_i \text{wait time})
\end{equation}

Assume we partition the time interval into $n$ sub-intervals, and there are $m$ GPUs are available. Let $T_F$ and $T_G$ denote the computation time of the fine and coarse solver on one partitioned sub-interval respectively. Let $T$ denote the total computation time of the fine solver run in serial, and $r>1$ the time cost ratio of the fine solver to the coarse solver. Then we have $T_F=\frac{T}{n}, T_G=\frac{T_F}{r}$. Let $l$ denote the number of iterations. For the regular scheduling strategy, the coarse phase of the $k$-th iteration needs to sequentially propagate across $n-k$ time intervals ($k=0,\ldots,l-1$).
During this stage, only one GPU is active, while the remaining $m-1$ GPUs remain idle. Hence, the idle time generated in the $k$-th iteration can be approximated as
\begin{equation}\label{eq:reg}
W_{\mathrm{reg}}^{(k)}\approx (m-1)(n-k)T_G.
\end{equation}
Summing over all iterations yields
\begin{equation}\label{eq:reg}
W_{\mathrm{reg}}\approx (m-1)T_G\sum_{k=0}^{l-1}(n-k).
=(m-1)T_G\Bigl(ln-\frac{l(l-1)}{2}\Bigr).
\end{equation}
For the pipelined scheduling strategy, the idle time mainly occurs during the initial injection stage.
The parallelism increases gradually from $1$ GPU to $m$ GPUs, and for $n\gg m$, the system remains nearly fully utilized after the pipeline is filled. Hence, the idle time can be approximated by
\begin{equation}\label{eq:pipe}
W_{\mathrm{pipe}}\approx \sum_{q=1}^{m-1}(m-q)T_G=\frac{m(m-1)}{2}T_G.
\end{equation}
By Equation \eqref{eq:reg} and \eqref{eq:pipe}, the difference in the total wait time between the total scheduling strategies is given by
\begin{equation}
\Delta W=W_{\mathrm{reg}}-W_{\mathrm{pipe}}\approx (m-1)T_G\Bigl(ln-\frac{l(l-1)}{2}-\frac{m}{2}\Bigr).
\end{equation}
Substituting $T_G=T_F/r$ gives
\begin{equation}\label{eq:delta_w}
\Delta W
\approx
(m-1)\frac{T}{rn}\Bigl(ln-\frac{l(l-1)}{2}-\frac{m}{2}\Bigr).
\end{equation}

Based on Equation \eqref{eq:delta_w}, we analyze the influence of $r$, $m$ and $n$.

\textbf{(1) Influence of $r$.}
The idle-time difference is proportional to $1/r$, i.e. $\Delta W \propto \frac{1}{r}$. A larger $r$ implies faster coarse propagation and less synchronization-induced idle time. However, the range of $r$ has an upper bound to ensure numerical stability. An excessively large $r$ may lead to loss of stability for stiff problems.

\textbf{(2) Influence of $m$.}
Let $A:=ln-\frac{l(l-1)}{2}$. Then $\Delta W\approx(m-1)\frac{T}{rn}\Bigl(A-\frac{m}{2}\Bigr)$. Treating $m$ as a continuous variable yields $\frac{d\Delta W}{dm}\propto A-m+\frac12$, indicating that $\Delta W$ reaches its maximum near $m^\ast \approx A+\frac12$. In practice, the effective time-parallelism is limited by the number of time intervals $n$. When $m\ge n$, adding more GPUs no longer increases parallelism and the performance gain saturates.

\textbf{(3) Influence of $n$.}
For fixed $T$ and $l$, $\Delta W=(m-1)\frac{T}{r}\Bigl(l-\frac{l(l-1)+m}{2n}\Bigr)$. For sufficiently large $n$, $\Delta W \sim (m-1)\frac{T}{r}\,l$, which implies $\Delta W$ becomes relatively insensitive to $n$ and is mainly determined by $m$, $l$, and $r$.

\subsection{Implementation of the solver}
We now present the implementation of the solver, organized into three parts. First, we describe the design of the fine and coarse propagators, each consisting of four stages: initialization, strain-twist vector computation for force calculation, velocity computation, and time integration to update positions and orthonormal triads. To ensure data coherency, we adopt a staged kernel execution model. Second, we present a highly parallelized computational pipeline covering each component of the computation, including force and moment evaluation, linear and angular velocity updates, a GPU-optimized matrix square root routine, and time integration. Finally, we detail a thread-mapping strategy designed for the swimmer simulation.

\subsubsection{Staged kernel execution and data coherency}
Following Eq.~\eqref{eq:serial_correct}, the Parareal algorithm consists of two components, $G$ and $F$, which correspond to the coarse and fine propagators, respectively. The coarse solver is typically a low-order scheme used to compute a coarse but fast approximation of the solution. In this work, we employ the explicit Euler method, which is a single-step scheme. The fine solver is typically a high-order numerical scheme that provides a fine and accurate approximation. Here, we present the case where an $n$th-order Runge–-Kutta method is used as the fine solver. In the numerical experiments, we set $n=2$, corresponding to the second-order Runge–Kutta method. 

\begin{figure}[h]
    \centering
    \includegraphics[width=\linewidth]{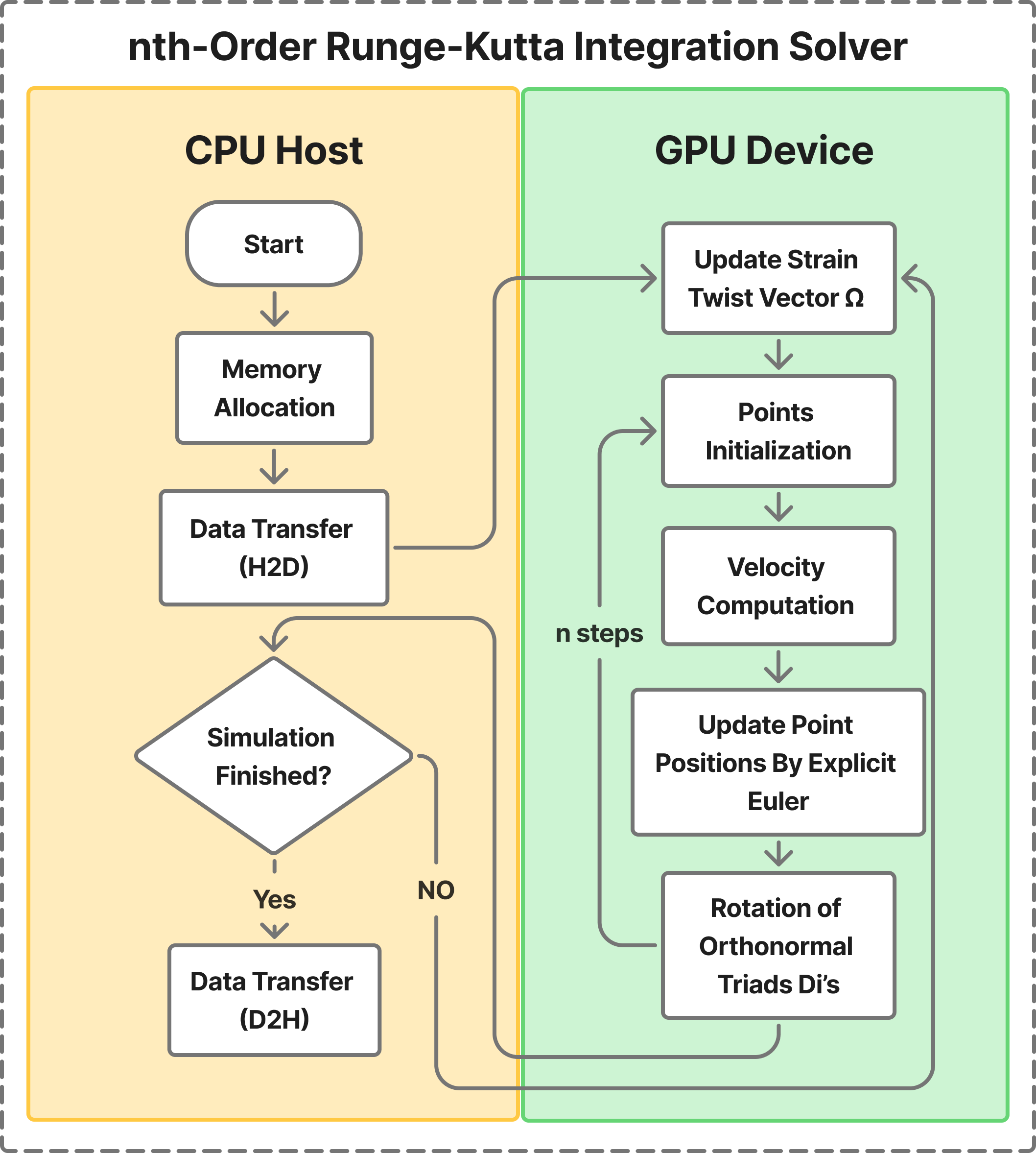}
    \caption{CPU--GPU coupled workflow for the nth-Order Runge--Kutta integration solver. The CPU handles global control and host--device data transfer, while the GPU executes the core computational kernels, including strain--twist evaluation, velocity computation, and explicit time integration. Arrows indicate execution flow and data dependencies.}
    \label{fig:CPU-GPU}
\end{figure}

In GPU-based rod simulations, CPU–-GPU data transfer frequently becomes a bottleneck when managed inefficiently. To minimize this overhead, we structure the solver using a staged execution pattern. Each time step is divided into four stages: initialization, force calculation, velocity evaluation, and configuration update. Each stage is implemented as an independent GPU kernel. To eliminate redundant data movement, all kernels share a consistent memory layout and operate on the same global memory region. Figure~\ref{fig:CPU-GPU} illustrates the flow of each solver. Arrows indicate data dependencies and execution order, while the layout shows how computation and data movement are organized in practice. By launching these kernels sequentially at each time step, our design keeps data resident on the device, reduces launch overhead, and enhances overall performance. Intermediate data are reused directly rather than being copied between kernels. This approach minimizes memory traffic and prevents frequent host–device synchronization. This strategy is depicted in Figure~\ref{fig:thread_map}, which will be discussed in more detail later.

\subsection{High-intensity parallel computational pipeline}

We now detail the kernel design, covering force and moment calculation, velocity calculation, and time integration. These kernels are implemented with \texttt{numba.cuda}, offering fine-grained control over thread and block configurations while supporting asynchronous execution to overlap computation and communication, thereby minimizing latency. Based on the structure of the problem, the computations are decomposed into components with varying levels of parallelism. This hierarchical mapping ensures each task is efficiently executed on the SIMT architecture, maximizing hardware utilization across all simulation scales.

\subsubsection{Local computation of forces and moments}\label{ssec:local_fn}

The discretization of $\bm{f}_{k}$ and $\bm{n}_{k}$ in Eqs.~\eqref{eq:discrete_f}--\eqref{eq:discrete_n} suggests that the force and moment at the $k$-th point can be written as
\begin{align}
    \bm{f}_{k} &= f(\bm{X}_{k-1},\bm{X}_{k},\bm{X}_{k+1}), \label{eq:force_eq} \\
    \bm{n}_{k} &= f(\bm{X}_{k-1},\bm{X}_{k},\bm{X}_{k+1},\bm{D}_{k-1},\bm{D}_{k},\bm{D}_{k+1}),\label{eq:torque_eq}
\end{align}
which shows that each point depends only on its immediate neighbors. As a result, the computation can be parallelized across discretization points.
\begin{algorithm}[H]
\caption{Local evaluation of internal force $\bm f_k$ and moment $\bm n_k$}
\label{alg:local_force_moment}
\begin{algorithmic}[1]

\ForAll{$k = 0,\dots,M-1$ \textbf{(in parallel)}}

\If{$0 < k < M-1$}

\State $\Delta \bm X_{k+\frac12} \gets \bm X_{k+1}-\bm X_k$
\State $\Delta \bm X_{k-\frac12} \gets \bm X_k-\bm X_{k-1}$

\State $\bm A_{k+\frac12} \gets \sum_{i=1}^3 \bm D_{i,k+1}\otimes\bm D_{i,k}$
\State $\bm S_{k+\frac12} \gets \sqrt{\bm A_{k+\frac12}}$

\State $\bm D^h_{i,k} \gets \bm S_{k+\frac12}\bm D_{i,k}$

\State Compute $\bm F_{k\pm\frac12}$ and $\bm N_{k\pm\frac12}$ using Eqs\eqref{eq:F_comp_discrete} and \eqref{eq:N_comp_discrete}.

\Else

\State Apply boundary conditions at $k=0$ or $k=M-1$

\EndIf
\EndFor

\end{algorithmic}
\end{algorithm}

A key difficulty in this step is the matrix square root in Eq.~\eqref{eq:mid_triad}. The standard CPU implementation, \texttt{scipy.linalg.sqrtm}, relies on general-purpose algorithms and is not suitable for GPU execution. In our case, however, the matrices involved are $3\times3$ rotation matrices. We exploit this structure by applying a specialized square-root algorithm for this problem \cite{Higham2008}.

For a rotation matrix $R$ with $\det R = 1$, there exists a unit vector $\mathbf{n} = (n_x,n_y,n_z)^\top$ and an angle $\theta$ such that
\begin{equation}
R = I\cos\theta + (1 - \cos\theta)\,\mathbf{n}\mathbf{n}^\top + \sin\theta\,K(\mathbf{n}),
\label{eq:rodrigues}
\end{equation}
where
\begin{equation}
K(\mathbf{n}) =
\begin{bmatrix}
0 & -n_z & n_y\\
n_z & 0 & -n_x\\
-n_y & n_x & 0
\end{bmatrix},
\quad
K(\mathbf{n})^2 = \mathbf{n}\mathbf{n}^\top - I.
\end{equation}
Equation~\eqref{eq:rodrigues} is the Rodrigues rotation formula.
Taking the trace yields $\cos\theta = \frac{\operatorname{tr}(R)-1}{2}$. If $\sin\theta\neq 0$, the rotation axis $\mathbf{n}$ can be determined from the skew-symmetric part:
\begin{equation}
\mathbf{n} = \frac{1}{2\sin\theta}
\begin{bmatrix}
R_{32}-R_{23}\\
R_{13}-R_{31}\\
R_{21}-R_{12}
\end{bmatrix}.
\end{equation}

Let a rotation matrix $S$ satisfy $S^2 = R$, and assume that it rotates about the same axis $\mathbf{n}$ by the half-angle $\theta/2$.
Then
\begin{equation}
S = I\cos\frac{\theta}{2} + 
\left(1 - \cos\frac{\theta}{2}\right)\mathbf{n}\mathbf{n}^\top +
\sin\frac{\theta}{2}\,K(\mathbf{n}).
\label{eq:halfangle}
\end{equation}
Using $K(\mathbf{n})^2 = \mathbf{n}\mathbf{n}^\top - I$ and
$K(\mathbf{n})\mathbf{n} = 0$, we obtain $S^2 = R$, which shows that Eq.~\eqref{eq:halfangle} provides a closed-form expression for the square root of a rotation matrix. In numerical implementations, instabilities may arise near the endpoints; we handle them as follows. When $\theta \approx 0$, we take $S \approx I$. When $\theta \approx \pi$, $\sin\theta$ becomes very small, and $\mathbf{n}$ should be estimated from the diagonal entries of $R$:
\begin{equation}
n_x^2 = \tfrac{1}{2}(R_{11}+1),\quad n_y^2 = \tfrac{1}{2}(R_{22}+1),\quad n_z^2 = \tfrac{1}{2}(R_{33}+1),
\end{equation}
and then set $S = R(\pi/2,\mathbf{n})$.

Compared to Schur-based methods, which involve iterative QR steps and complex control flow, the proposed approach uses fixed-size matrix operations and a predictable sequence of instructions. This makes it more suitable for GPU execution. In practice, we observe that this implementation is both faster and sufficiently accurate compared with \texttt{scipy.linalg.sqrtm}.

\subsubsection{Computation of linear and angular velocities}\label{ssec:vel}
Based on Equation \eqref{eq:superposition}, the linear and angular velocities can be written in the form:
\begin{align}        
\bm u &=\frac{1}{\mu}\bigl(Q_{ij}\bm f_j + P_{ij}\bm n_j\bigr)\\
\bm\omega&=\frac{1}{\mu}(\mathcal{W}_{ij}\bm f_j + R_{ij}\bm n_j)
\end{align}
where $Q_{ij}(\varepsilon, \bm{r}_{ij})$, $P_{ij}(\varepsilon, \bm{r}_{ij})$, $\mathcal{W}_{ij}(\varepsilon, \bm{r}_{ij})$ and $R_{ij}(\varepsilon, \bm{r}_{ij})$ are matrices resulting from the regularized kernel functions associated with MRS for $\bm r_{ij}=\bm x_i-\bm x_j$. Eq.~\eqref{eq:superposition} implies that the linear and angular velocities at all points on the rod are independent during evaluation, meaning the summation order can be arbitrarily permuted. This property allows the velocity computation to be parallelized. The GPU-based parallelization is presented in Algorithm \ref{alg:vel_shared_simple}.

\begin{figure}[h]
    \centering
    \includegraphics[width=\linewidth]{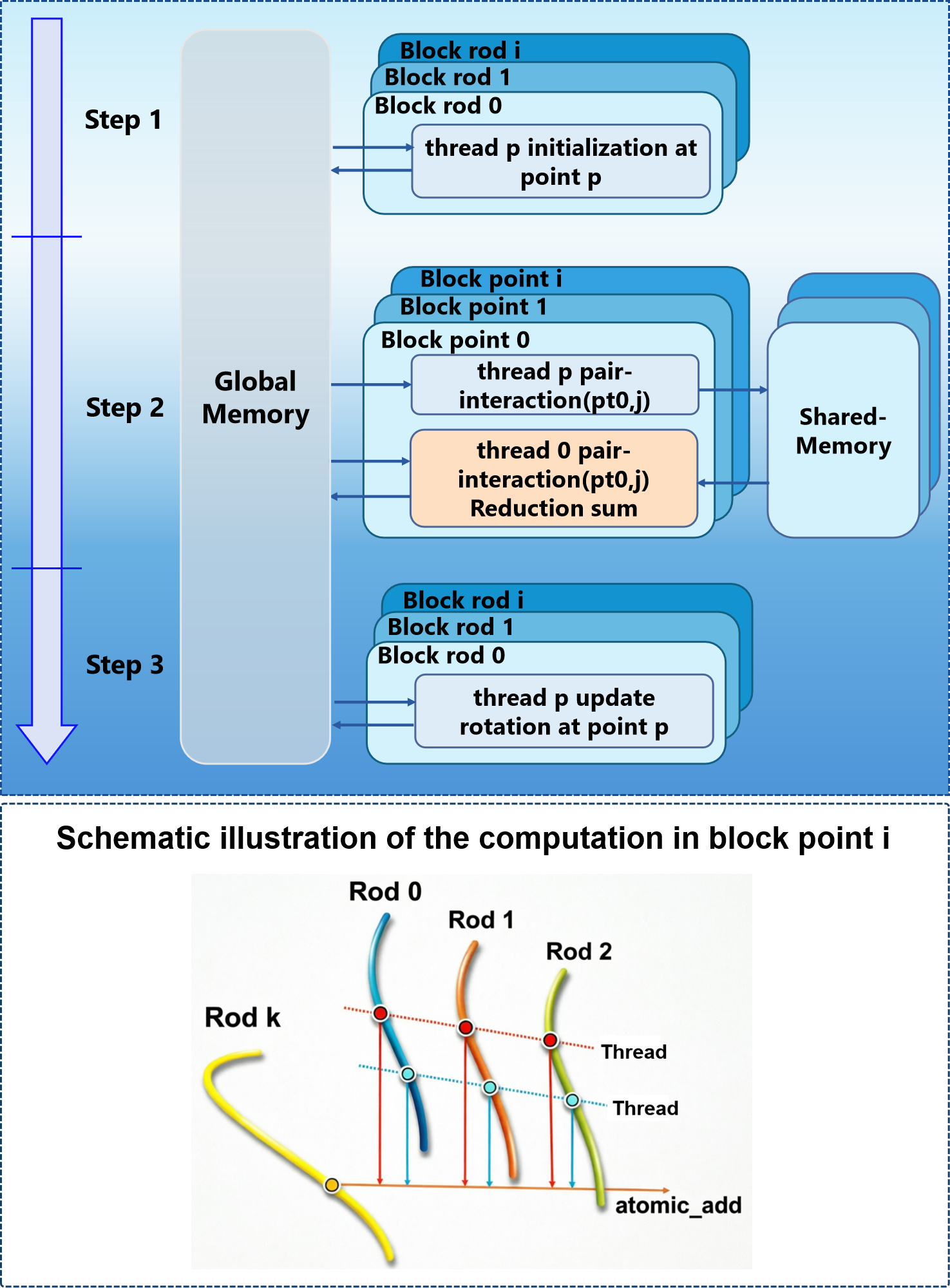}
    \caption{Thread--block mapping and data flow in the GPU implementation. 
    In Step~2, each thread $p$ in block $i$ processes a subset of source indices in a strided manner, i.e., $j = p,\, p + T,\, p + 2T, \ldots$, where $T$ is the number of threads per block. 
    The lower panel illustrates the interaction pattern at a target point, where threads accumulate contributions from multiple rods.}
    \label{fig:thread_map}
\end{figure}
\subsubsection{Time integration}
In the temporal dimension, we update the rod position using the explicit Euler scheme. Accordingly, the position of the $k$-th point on the rod at time step $n+1$ can be written as
\begin{equation}
\bm{X}_k^{n+1} = \bm{X}_k^n +\bm{u}(\bm{X}_k^n) \Delta t,
\label{eq:newx}
\end{equation}
The angular velocity $\boldsymbol{\omega}_k^n = \boldsymbol{\omega}(\bm{X}_k^n)$ is computed using Eq.~\eqref{eq:superposition}, and the orthonormal triads $\bm{D}_i$'s $(i=1,2,3)$ are updated using Rodrigues' rotation formula:
\begin{equation}
\begin{aligned}
(\bm{D}_k^{i})^{n+1} = \cos\theta \, (\bm{D}_k^{i})^{n} &+ \sin\theta \cdot \bigl(\bm{e} \times (\bm{D}_k^{i})^{n}\bigr) \\
&+ (1 - \cos\theta) \bigl(\bm{e} \cdot (\bm{D}_k^{i})^{n}\bigr) \bm{e},
\label{eq:newD}
\end{aligned}
\end{equation}
where $\theta = |\boldsymbol{\omega}_k^n| \Delta t$, $\bm{e} = \boldsymbol{\omega}_k^n / |\boldsymbol{\omega}_k^n|$ is the unit vector along the rotation axis, ``$\times$'' denotes the cross product, and ``$\cdot$'' denotes the dot product. 

\begin{algorithm}[h]
\caption{Parallel evaluation of $\bm u(\bm x_i)$ and $\bm\omega(\bm x_i)$ using shared-memory reduction}
\label{alg:vel_shared_simple}
\begin{algorithmic}[1]
\Require Targets $\{\bm x_i\}_{i=1}^{N}$; sources $\{\bm X_j,\bm f_j,\bm n_j\}_{j=1}^{N}$;
regularization parameter $\varepsilon$; viscosity $\mu$; (optional) wall and LJ parameters.
\Ensure $\bm u(\bm x_i)$ and $\bm\omega(\bm x_i)$ for all $i$.

\ForAll{$i=1,\dots,N$ \textbf{in parallel}} \Comment{one CUDA block per target point}
    \State $\bm u \gets \bm 0,\ \bm\omega \gets \bm 0$ stored in \textbf{shared memory}
    \State \textbf{syncthreads}

    \ForAll{$j=1,\dots,N$ distributed among threads} \Comment{each thread processes a subset of sources}
        \State Compute $\bm r_{ij} \gets \bm x_i - \bm X_j$, $r_{ij}=\|\bm r_{ij}\|$
        \State Evaluate kernels $Q_{ij}(\varepsilon,\bm r_{ij})$ and $P_{ij}(\varepsilon,\bm r_{ij})$
              \Comment{include wall/image correction if needed}
        \State (optional) modify $\bm f_j \leftarrow \bm f_j + \bm f^{\mathrm{LJ}}_{ij}$ if LJ is enabled
        \State $\Delta \bm u \gets \frac{1}{\mu}\bigl(Q_{ij}\bm f_j + P_{ij}\bm n_j\bigr)$
        \State $\Delta \bm\omega \gets \frac{1}{\mu}\,(\mathcal{W}_{ij}\bm f_j + R_ij\bm n_j)$
              \Comment{based on Eqs.~\eqref{eq:velocity_field}--\eqref{eq:angular_velocity_field}}
        \State \textbf{atomicAdd}$(\bm u,\Delta\bm u)$ in shared memory
        \State \textbf{atomicAdd}$(\bm\omega,\Delta\bm\omega)$ in shared memory
    \EndFor

    \State \textbf{syncthreads}
    \State Write $\bm U(:,i)\gets \bm u$ and $\bm W(:,i)\gets \bm\omega$ to global memory (one thread)
\EndFor
\end{algorithmic}
\end{algorithm}

\subsubsection{Thread mapping strategy}

The various kernels within the solver exhibit distinct data dependencies and computational patterns, each requiring a tailored thread-block mapping strategy. To address this, we design such a strategy. For local computations, such as internal force and moment evaluation, we map each discretization point to an individual thread and each rod to a dedicated thread block. Since each point depends only on its immediate neighbors, this mapping preserves spatial data locality and enables efficient reuse of rod-level data within high-speed shared memory.

In contrast, nonlocal computations, specifically the velocity evaluation based on MRS, employ a target-based mapping strategy. Each target point is assigned to a thread block, within which threads iterate over the corresponding source points. This decomposition exposes the inherent parallelism in source–target point interactions, allowing the GPU to process multiple interactions concurrently while accumulating partial results in shared memory to minimize global memory access.

For time integration and other element-wise updates, each discretization point is handled by a single thread. This stage is embarrassingly parallel and requires no synchronization between threads. Collectively, these mapping strategies are meticulously aligned with the mathematical structure of each computational task, ensuring high parallel efficiency and optimal GPU resource utilization. A schematic representation of the thread mapping strategy is shown in Figure~\ref{fig:thread_map}.

\section{Numerical Results}
In this section, we apply the proposed framework to the simulations of filamentous microswimmers in viscous fluid. We first verify numerical convergence and report the solution error. We then evaluate the matrix square root algorithm, demonstrating the superiority of the GPU-optimized routine. Finally, we present scaling test results, which also validate the asymptotic analysis from Section~\ref{ssec:cost}.

\subsection{Environment Setup}
All experiments are implemented in Python. The GPU platform is equipped with an NVIDIA A100 PCIe 40GB GPU running on an aarch64 architecture with Kylin Linux Advanced Server V10. The CPU is a Kunpeng-920 processor at 3.0\,GHz with 220\,GB of available memory. The CPU-only experiments are performed on an AMD 7H12 processor (2.6\,GHz) with 128 cores and 512\,GB memory. Unless otherwise specified, the physical parameters remain identical across all experiments. 

We model a rod-shaped filament swimmer with the Kirchhoff rod formulation, a common approach used in computational studies of flagellar swimmers \cite{carichino2019emergent,olson2013modeling,Lim2010,Olson2014}. The rod's centerline, is a space curve that defines its geometry, while its flagellar motion follows a planar sinusoidal wave, modeled based on experimental observations \cite{Ho2001HyperactivationOM,Smith2009BendPI}. As in \cite{olson2013modeling}, the waveform is imposed via a prescribed preferred strain–twist vector:
\begin{equation}
\left(\Omega_1,\Omega_2,\Omega_3\right)=\left(0,-k^2A\sin(ks+f t),0\right),
\end{equation}
where $s$ denotes arclength, $A$ the amplitude, $f$ the frequency, and $k=2\pi/\lambda$ with $\lambda$ the wavelength. The rod is situated in a semi-infinite fluid domain $\{(x,y,z)\in\mathbb{R}^3 \mid z \ge 0\}$, bounded by a stationary, infinite planar wall at $z=0$ where the no-slip condition holds. Initially, at $t=0$, the rod is straight and parallel to the $z$-plane, positioned at a height $d_z=1$ above the wall. Each rod swimmer is initialized as straight and placed in the domain with random orientation and position. Each rod is discretized into $M$ points along the rod length, with $\Delta s=L/(M-1)$. In the following numerical experiments, we set $M=51$ and the regularization parameter of the MRS method to $\epsilon=4\Delta s$. To prevent contact between swimmers, a repulsive force based on the Lennard-Jones potential $U(r) = 4\varepsilon \left[ \left(\frac{\sigma}{r}\right)^{12} - \left(\frac{\sigma}{r}\right)^{6} \right]$ is applied when the pairwise distance $r$ between two points falls below a threshold $r_c=2^{1/6}\sigma$, with $\sigma=3\epsilon$. The time step size used in the fine solver is fixed at $\Delta t = 10^{-6}$. 

\begin{table*}[t]
\caption{Runtime comparison of CPU and GPU implementations for different solver components.}
\centering
\setlength{\tabcolsep}{4pt}
\small
\resizebox{\linewidth}{!}{
\begin{tabular}{c|ccc|ccc|ccc|ccc|c}
\hline
\multirow{2}{*}{Rods}
& \multicolumn{3}{c|}{Initialization}
& \multicolumn{3}{c|}{Velocity computation}
& \multicolumn{3}{c|}{Orthonormal triads' update}
& \multicolumn{3}{c|}{Total}
& $T_{\mathrm{comm}}$ \\
\cline{2-14}
& CPU & GPU & Speedup
& CPU & GPU & Speedup
& CPU & GPU & Speedup
& CPU & GPU & Speedup
&  \\
\hline
1  & 0.010812 & 0.000683 & 15.83$\times$
   & 0.013908 & 0.000389 & 35.79$\times$
   & 0.006234 & 0.000230 & 27.15$\times$
   & 0.030954 & 0.001301 & 23.80$\times$
   & 0.00373 \\

4  & 0.044339 & 0.000680 & 65.20$\times$
   & 0.073934 & 0.000381 & 193.97$\times$
   & 0.024397 & 0.000229 & 106.56$\times$
   & 0.142669 & 0.001290 & 110.60$\times$
   & 0.00376 \\

12 & 0.132558 & 0.000676 & 196.05$\times$
   & 0.308140 & 0.000382 & 805.78$\times$
   & 0.073840 & 0.000230 & 321.33$\times$
   & 0.514539 & 0.001288 & 399.52$\times$
   & 0.00401 \\

25 & 0.276209 & 0.000687 & 402.03$\times$
   & 0.887801 & 0.000386 & 2301.74$\times$
   & 0.152188 & 0.000638 & 238.60$\times$
   & 1.316198 & 0.001711 & 769.29$\times$
   & 0.22368 \\
\hline
\end{tabular}}

\label{tab:A_kernel_speedup}
\end{table*}
\subsection{Convergence Verification}
To evaluate the accuracy and verify the convergence of the solver, we examine two metrics: the true relative error $\eta^k$, which measures the error with respect to the true solution, and the relative increment $\widetilde{\eta}^k$, which quantifies the difference between consecutive iterations obtained at the $k$th Parareal iteration.
\begin{align}\label{eq:true_err}
\eta^k &=
\max_{i=1,~2,~\cdots,~N_{\sigma}}\frac{\left\|\mathbf{x}_i^k-\mathbf{x}_i^{\mathcal{F}}\right\|_2}{\left\|\mathbf{x}_i^{\mathcal{F}}\right\|_2} \\
\widetilde{\eta}^k &=
\max_{i=1,~2,~\cdots,~N_{\sigma}}\frac{\left\|\mathbf{x}^k_i-\mathbf{x}^{k-1}_i\right\|_2}{\left\|\mathbf{x}^k_i\right\|_2}~~\text{for}~~k=1,~2,~\cdots
\end{align}
Here, $\mathbf{x}_i^k$ is the position of the $i$th point at iteration $k$, $\mathbf{x}i^{\mathcal{F}}$ is the fully serial fine-solver solution, and $N{\sigma}$ is the number of grid points. Because the serial solution is not known in practice, the stopping criterion instead uses the relative increment, i.e. the difference between successive Parareal iterates.

\begin{figure}[H]
\centering
\begin{subfigure}{0.48\linewidth}
\centering
\includegraphics[width=\linewidth]{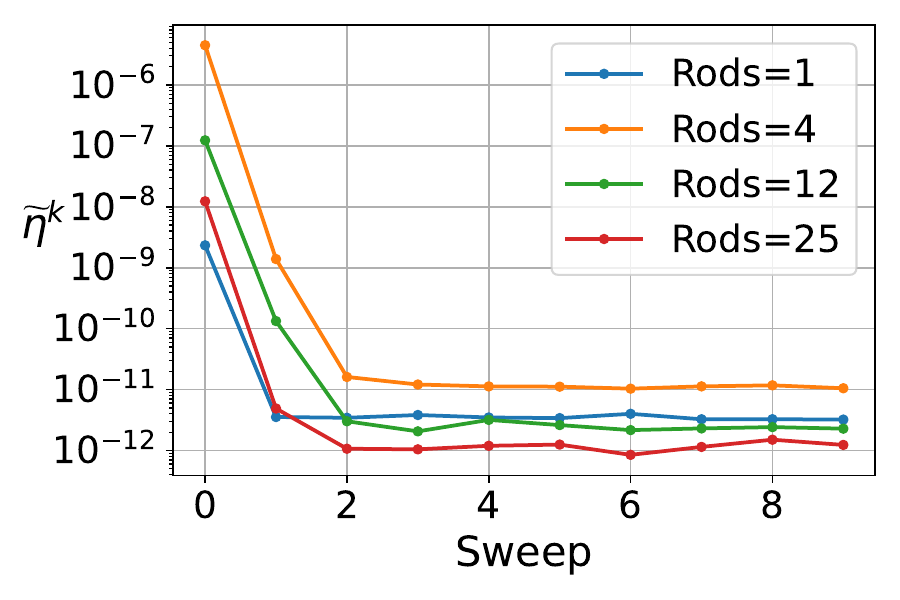}
\caption{}
\end{subfigure}
\begin{subfigure}{0.48\linewidth}
\centering
\includegraphics[width=\linewidth]{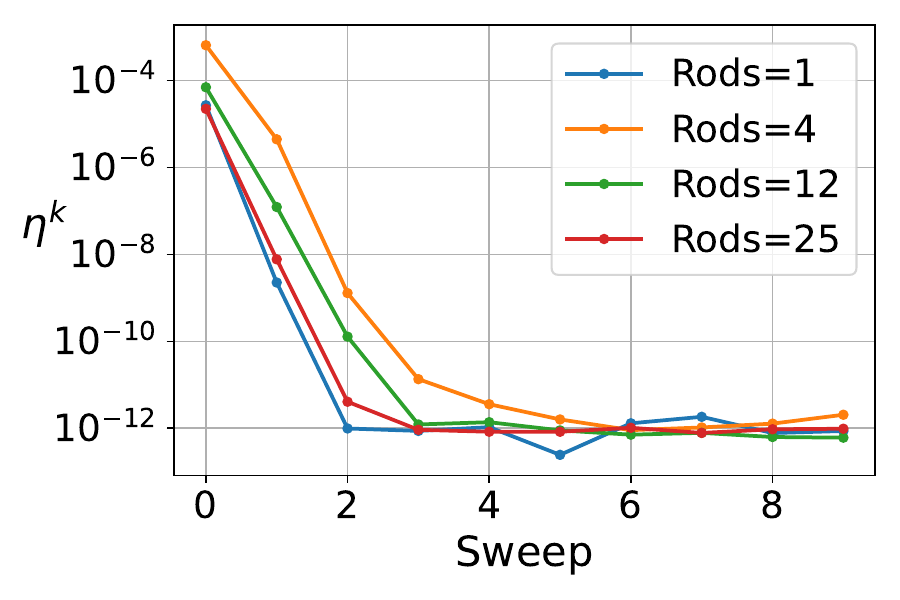}
\caption{}
\end{subfigure}
\caption{(a) Relative increment $\widetilde{\eta}^k$ (b) True relative error $\eta^k$}
\label{fig:combined_convergence}
\end{figure}

We first verify the convergence of the time-parallel algorithm. The time-parallel solution is compared with both the reference serial solution and the previous iteration. We solve for the filament dynamics for $t\in(0,T]$ with $T=1$ with number of rods $1,4,12,25$. Fig.~\ref{fig:combined_convergence} (a) presents the relative increment $\widetilde{\eta}^k$ and Fig.~\ref{fig:combined_convergence} (b) shows the true relative error $\eta^k$ obtained in each simulation. Both quantities are plotted on a semi-logarithmic scale. The results show that the error decay steadily and that approximately four iterations suffice for a solution accuracy $\eta^k$ of order $10^{-12}$, indicating good convergence of the algorithm.

\subsection{GPU Spatial Parallel Performance}
First, we examine the performance gain of spatial parallelization on a single GPU. We perform experiments for the rod count $1,4,12,25$. In each case, we measure the time cost of the three main components that make up each time step's calculation: the initialization, velocity computation, and orthonormal triads' update. To evaluate the relative performance improvement of the proposed GPU-based framework to the CPU implementation, we calculate the relative speedup defined as the ratio between the execution time of the CPU implementation and that of the GPU implementation 
\begin{equation}
S_1 = \frac{T_{\mathrm{CPU}}}{T_{\mathrm{GPU}}}.
\label{eq:speedup_spatial}
\end{equation}
We also measure the communication cost, which is the total data transfer time between the host and the GPU device, i.e. $T_{\mathrm{comm}} = T_{\mathrm{H2D}} + T_{\mathrm{D2H}}$, where $T_{\mathrm{H2D}}$ and $T_{\mathrm{D2H}}$ denote the host-to-device and device-to-host transfer times illustrated in Figure~\ref{fig:CPU-GPU}, accumulated over a single solver invocation. In Table~\ref{tab:A_kernel_speedup}, we report the runtime and corresponding speedup of the CPU and GPU implementations for each component.

We observe that the CPU runtime increases approximately linearly with the number of rods, with the velocity computation dominating the overall computational cost. In contrast, the GPU runtime remains nearly constant from 1 to 12 rods, indicating that spatial parallelism effectively absorbs the increased workload. When the number of rods increases to 25, the update of orthonormal triads' shows a slight increase in runtime, but still remains significantly faster than its CPU counterpart. In this case, the overall speedup reaches $769\times$, demonstrating the effectiveness of spatial GPU parallelization.

\subsection{Performance and Accuracy of the Matrix Square Root Implementation}
To evaluate the performance and numerical stability of the CUDA implementation of the $3 \times 3$ matrix square root algorithm, we select a random unit vector as the rotation axis and uniformly sample 100 rotation angles $\theta \in [-0.1, \pi + 0.1]$ to generate rotation matrices. We compute the matrix square root of the rotation matrices using SciPy's \texttt{sqrtm} and the half-angle rotation routine described in Section \ref{ssec:local_fn}.

\begin{figure}[H]
\centering
\begin{subfigure}{0.48\linewidth}
\centering
\includegraphics[width=\linewidth]{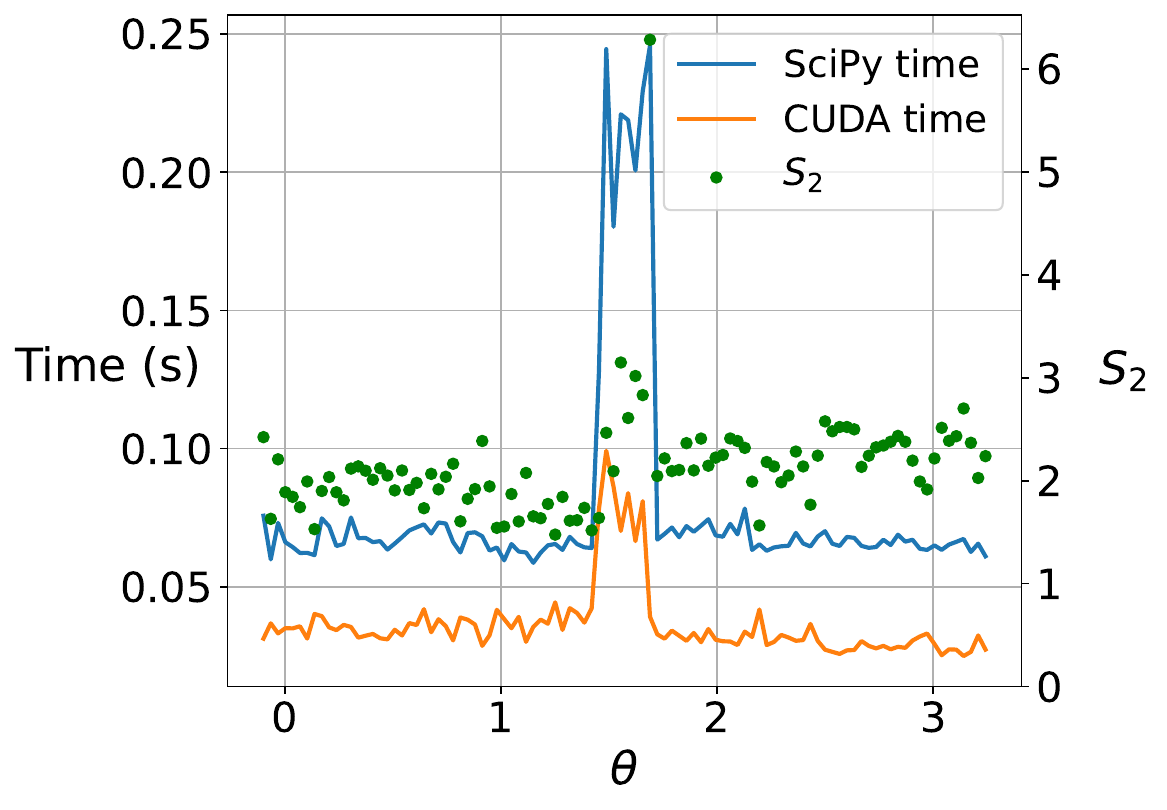}
\caption{}
\end{subfigure}
\begin{subfigure}{0.48\linewidth}
\centering
\includegraphics[width=\linewidth]{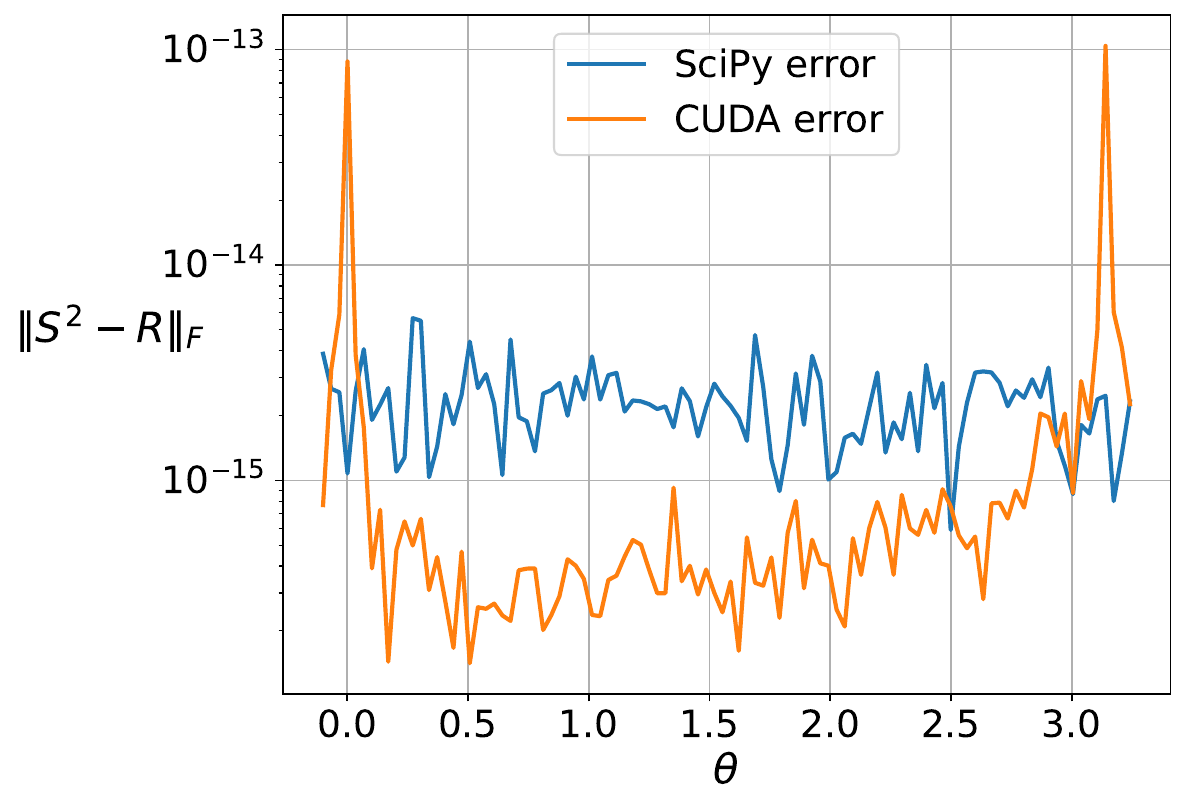}
\caption{}
\end{subfigure}
\caption{Performance and accuracy of the $3\times3$ matrix square root implementation: (a) speedup (b) residual}
\label{fig:sqrt_results}
\end{figure}

To demonstrate its improvement over SciPy's \texttt{sqrtm}, We calculate the speedup of the GPU-optimized routine relative to \texttt{sqrtm} $S_2 = \frac{T_{\mathrm{SciPy}}}{T_{\mathrm{CUDA}}}$, where $T_{\mathrm{SciPy}}$ and $T_{\mathrm{CUDA}}$ denote the execution times of of the two implementations respectively. In Fig.~\ref{fig:sqrt_results} (a), we present the speedup and the numerical error of each method. The CUDA implementation achieves approximately $2\times$ computation speed across most angles, with an average speedup of 2.14. We present the error $\|S^2-R\|_F$ in Fig.~\ref{fig:sqrt_results} (b), and the associated statistical error and time metrics in Table~\ref{tab:33stats}. As summarized in Table~\ref{tab:33stats}, the proposed method achieves comparable or higher accuracy than SciPy's \texttt{sqrtm}. Numerical instability may appear when $\theta\approx0,\pi$, which results in several outliers in the error distribution. This phenomenon increases the variance of the error but does not affect the convergence behavior of the overall algorithm.
\begin{table}[htbp]
\centering
\caption{Statistical summary of performance and numerical accuracy.}
\begin{tabular}{lccc}
\hline
Metric & Mean & Median & Std / Max \\
\hline
Speedup 
& $2.14$ 
& $2.10$ 
& $0.53$ \\

SciPy error 
& $2.33\times10^{-15}$ 
& $2.31\times10^{-15}$ 
& $9.68\times10^{-16}$ \\

CUDA error 
& $2.76\times10^{-15}$ 
& $4.71\times10^{-16}$ 
& $1.35\times10^{-14}$ \\

SciPy time  (s)
& $7.81\times10^{-2}$ 
& $6.63\times10^{-2}$ 
& $2.46\times10^{-1}$ \\

CUDA time (s)
& $3.64\times10^{-2}$ 
& $3.28\times10^{-2}$ 
& $9.92\times10^{-2}$ \\
\hline
\end{tabular}
\label{tab:33stats}
\end{table}

\begin{table*}[t]
\caption{Runtime comparison between the regular and pipelined scheduling.
Each entry reports \emph{regular / pipeline} runtime.}
\centering
\resizebox{\linewidth}{!}{
\begin{tabular}{c|ccc|ccc}
\hline
 & \multicolumn{3}{c|}{2 GPUs} & \multicolumn{3}{c}{4 GPUs} \\
Rods & $r=2$ & $r=5$ & $r=8$ & $r=4$ & $r=10$ & $r=16$ \\
\hline
1 &
815.71 / 552.87 &
508.06 / 408.64 &
413.56 / 381.23 &
810.22 / 548.23 &
501.01 / 420.98 &
424.12 / 381.63 \\

4 &
830.77 / 558.27 &
490.23 / 412.75 &
431.24 / 389.06 &
827.48 / 552.94 &
503.83 / 417.04 &
424.42 / 379.09 \\

12 &
1420.24 / 1109.08 &
1111.41 / 938.49 &
1031.39 / 898.03 &
970.64 / 729.38 &
664.74 / 572.98 &
590.75 / 536.67 \\

25 &
3461.85 / 2819.40 &
2756.28 / 2381.44 &
2582.23 / 2277.16 &
2333.84 / 1749.87 &
1623.55 / 1294.12 &
1447.81 / 1238.45 \\

\hline
\end{tabular}}

\label{tab:runtime_compare}
\end{table*}

\subsection{Time-Parallel Performance}
In this subsection, we perform filament simulations on multiple GPUs and compare the efficiency of the pipelined Parareal method with that of the regular Parareal method. Simulations are conducted for rod counts of 1, 4, 12, and 25. Let $r = T_F / T_G$. We run the simulations on 2 GPUs with $r = 2, 5, 8$, and on 4 GPUs with $r = 4, 10, 16$. Table~\ref{tab:runtime_compare} reports the runtime comparison between the regular time-parallel scheduling and the pipelined scheduling under different parameter settings. Each entry is presented in the format \emph{regular / pipeline}. Under all tested configurations, the pipelined scheduling consistently outperforms the regular scheduling, and this advantage holds across simulations involving different rod counts.

From the numerical results, it is also clear that the performance gap between the two scheduling strategies is more pronounced when $r$ is small. For instance, when $r = 2$ (or $r = 4$), the pipelined Parareal reduces runtime by approximately 25\%–-30\% compared to the regular Parareal. As $r$ increases, this gap gradually diminishes. This trend is consistent with the idle-time analysis presented in Section~\ref{ssec:cost}: when the coarse solver becomes faster (i.e., as $r$ increases), the coarse wavefront propagates more quickly, reducing the idle waiting time caused by synchronization barriers in the regular scheduling. Consequently, the runtime difference between the two scheduling strategies becomes smaller.

\begin{figure}[h]
\centering

\begin{minipage}{0.42\linewidth}
\centering
\includegraphics[width=\linewidth]{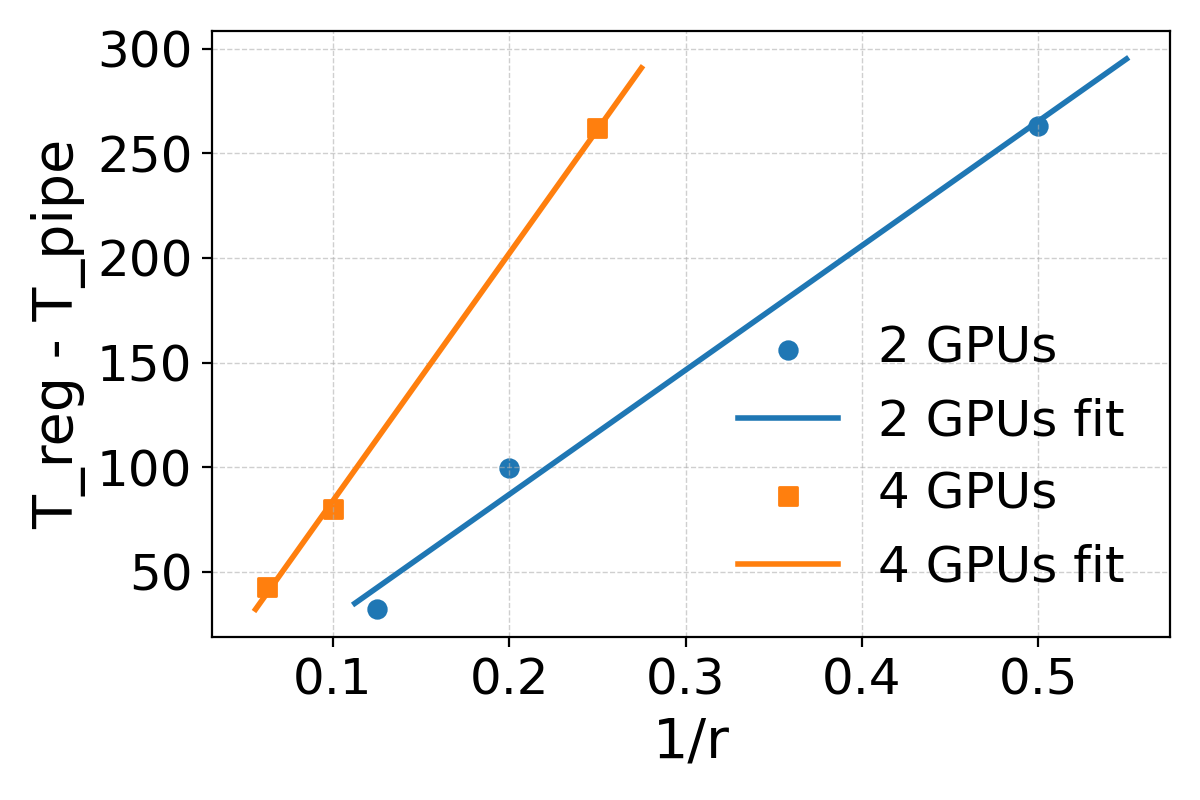}
\\ (a) 
\end{minipage}
\begin{minipage}{0.42\linewidth}
\centering
\includegraphics[width=\linewidth]{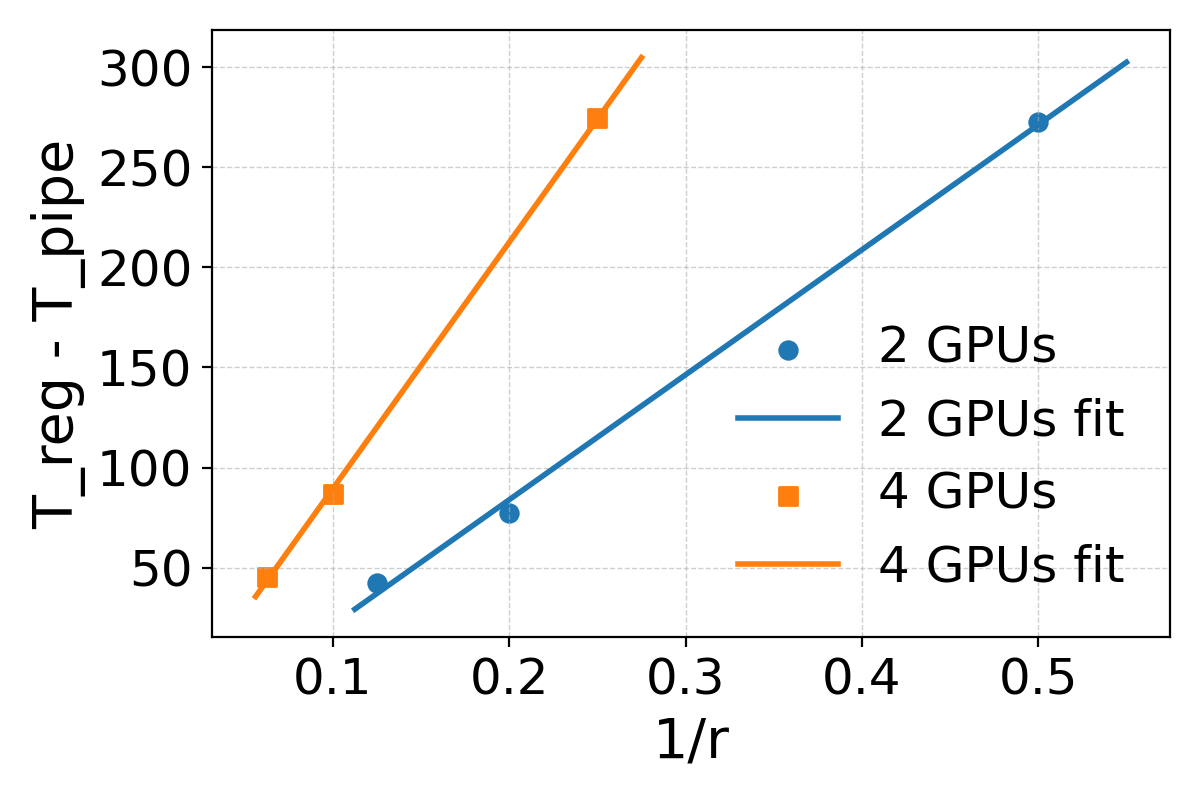}
\\ (b) 
\end{minipage}
\begin{minipage}{0.42\linewidth}
\centering
\includegraphics[width=\linewidth]{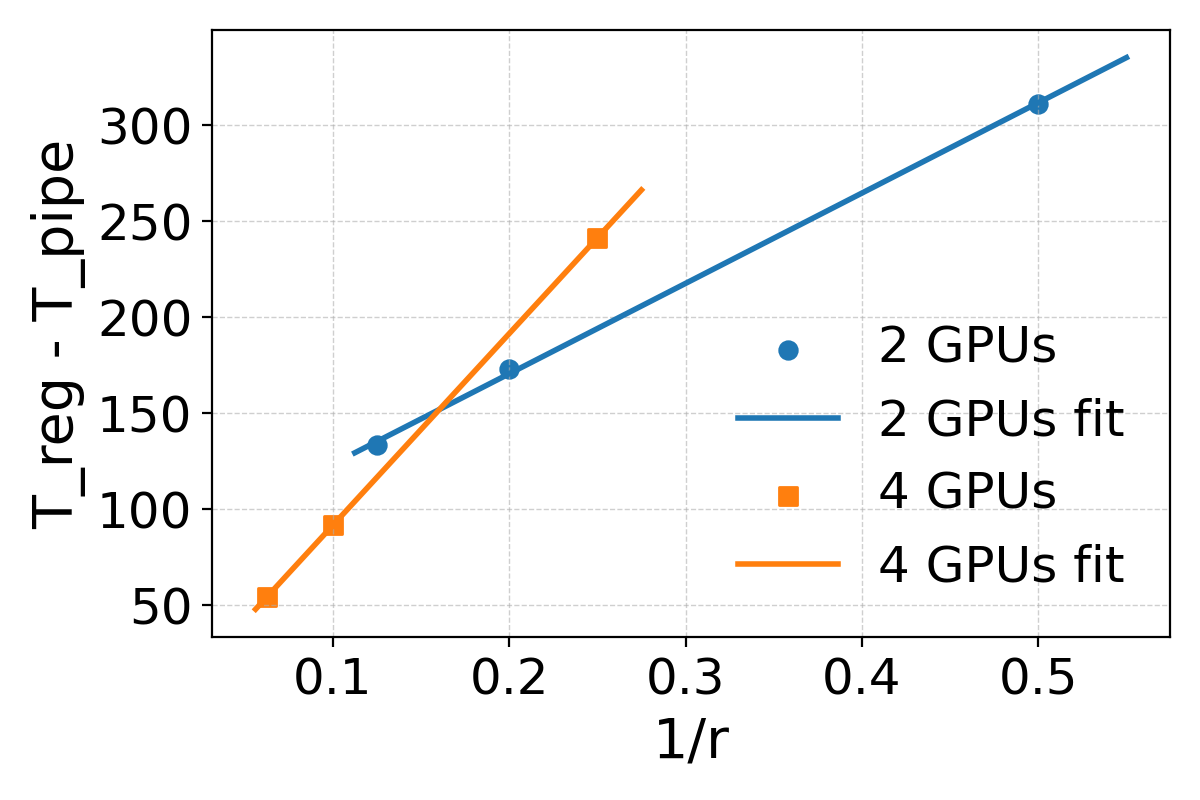}
\\ (c) 
\end{minipage}
\begin{minipage}{0.42\linewidth}
\centering
\includegraphics[width=\linewidth]{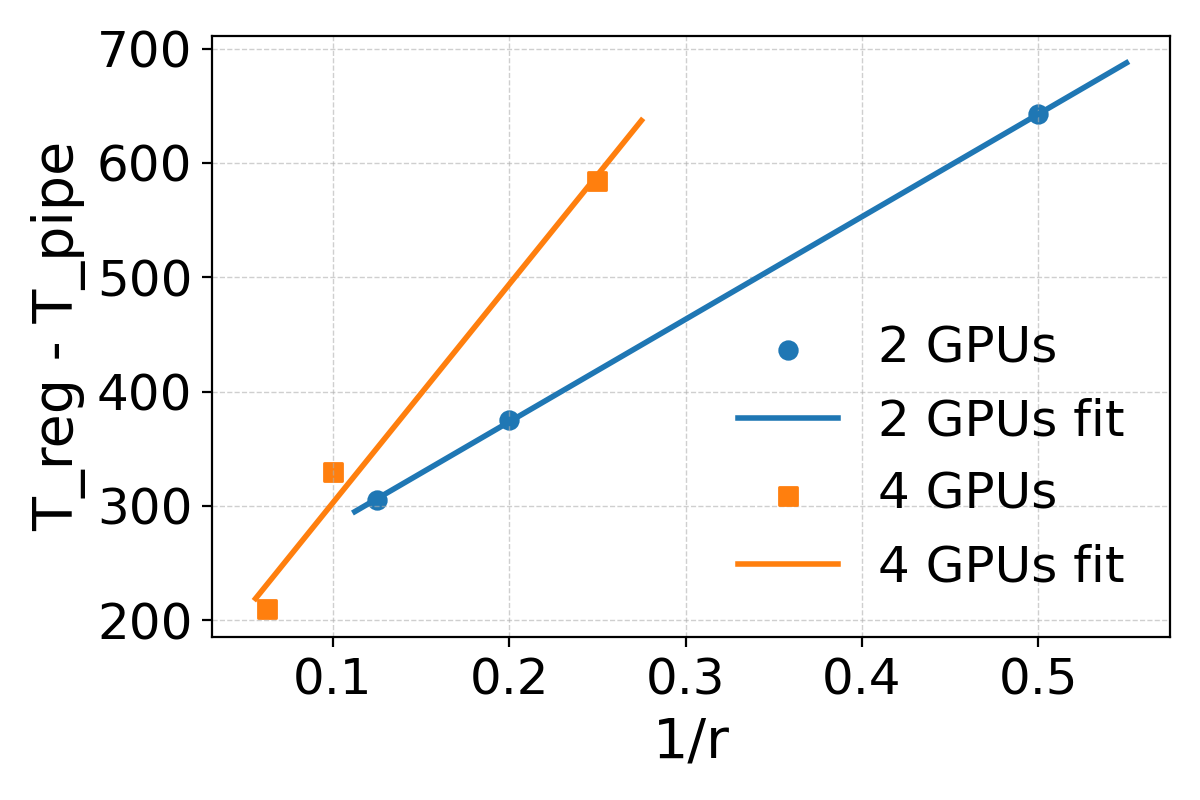}
\\ (d) 
\end{minipage}

\caption{Runtime gap $T_{\mathrm{reg}}-T_{\mathrm{pipe}}$ versus $1/r$ for (a) 1 rod (b) 4 rods (c) 12 rods (d) 25 rods.}
\label{fig:gap_inverse_r}
\end{figure}

To further validate the theoretical analysis presented in Section~\ref{ssec:cost}, we plot the measured runtime gap $T_{\mathrm{reg}}-T_{\mathrm{pipe}}$ vs. $1/r$ in Fig.~\ref{fig:gap_inverse_r}. For all rod counts considered in the experiments (1, 4, 12, and 25), the measured data exhibit an approximately linear trend, and the fitted lines agree well with the experimental points. This behavior provides direct empirical evidence for the theoretical prediction obtained from the GPU wait-time difference estimate $\Delta W$. Since the regular and pipeline scheduling strategies execute exactly the same sets of coarse and fine computations, their total computational workload is essentially identical. Hence, the dominant source of runtime difference $\Delta T=T_{\rm reg}-T_{\rm pipe}$ is the difference in GPU idle time. Under this assumption, it is reasonable to approximate $\Delta T \approx \Delta W$. The experimental observation $T_{\mathrm{reg}}-T_{\mathrm{pipe}} \propto \frac{1}{r}$ is therefore consistent with the theoretical prediction $\Delta W \propto \frac{1}{r}$. This agreement indicates that the measured total runtime gap is indeed primarily determined by the difference in idle waiting time, and thus provides a validation of the analysis of $\Delta W$ in GPU idle-time analysis. Based on the analysis, this advantage becomes more pronounced for longer simulations, i.e. larger $T$ in equation \eqref{eq:delta_w}.

In the regular Parareal scheduling, the additional idle time mainly arises from the $(m-1)$ GPUs waiting during the sequential coarse propagation. Hence, $\Delta T$ has a dominant term that scales with $(m-1)$, with an additional term proportional to $-m/2$, which accounts for the finite cost of establishing parallelism in the pipeline strategy. The experimental results reflect the same trend. For instance, in the case of 25 rods, when $m=2$ (with $r=2,5,8$), the linear fit of $\Delta T$ versus $1/r$ gives a slope of approximately $0.9\times10^3$; when $m=4$ (with $r=4,10,16$), the corresponding slope increases to approximately $2.0\times10^3$. The increase in slope with respect to $m$ indicates that a larger number of GPUs amplifies the synchronization-induced waiting cost in the regular scheduling. Although this growth is not strictly proportional to $(m-1)$ due to the $-m/2$ correction and additional non-idle overheads such as communication and memory access, the overall trend remains consistent with the theoretical prediction, that is, as the number of GPUs increases, the performance advantage of the pipeline scheduling over the regular scheduling becomes more pronounced.

\subsection{Weak Scaling}
\begin{figure}[H]
\centering
\begin{subfigure}{0.48\linewidth}
\centering
\includegraphics[width=\linewidth]{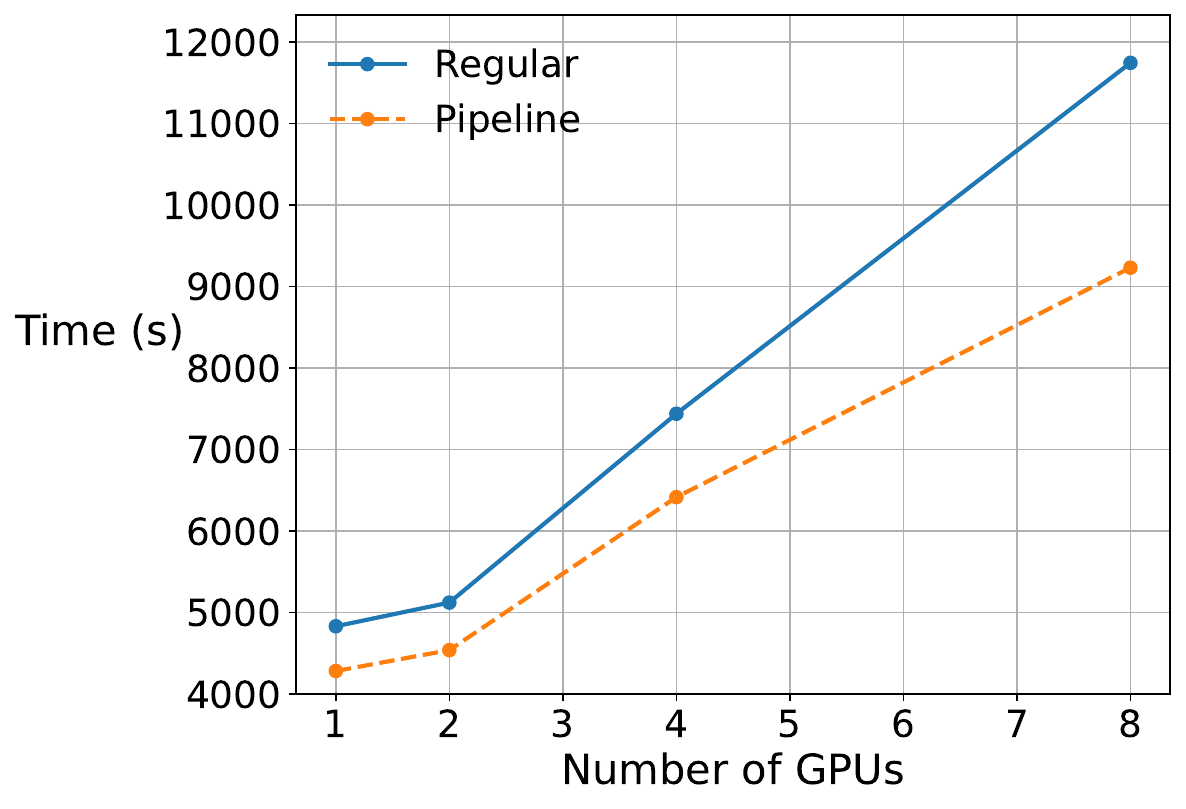}
\caption{}
\end{subfigure}
\begin{subfigure}{0.48\linewidth}
\centering
\includegraphics[width=\linewidth]{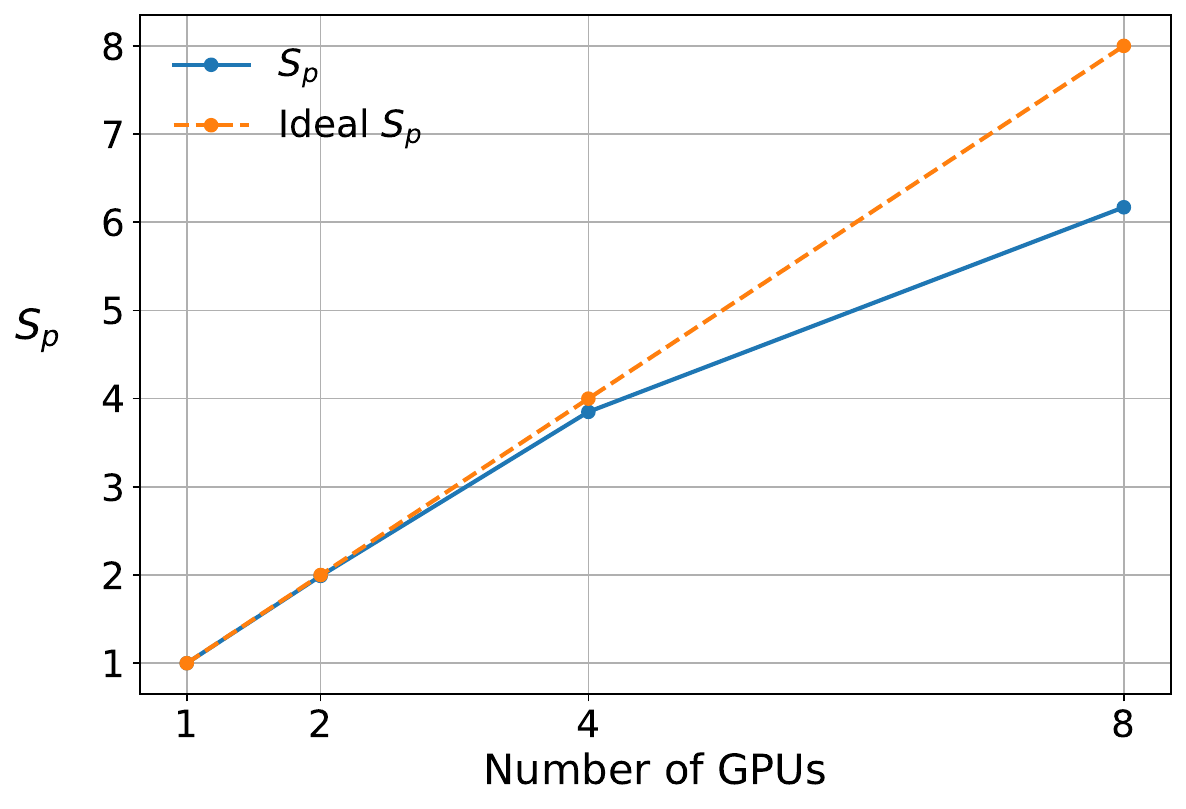}
\caption{}
\end{subfigure}
\caption{Scalability results of the proposed solver:
(a) Weak scaling performance as the problem size and GPU number increase proportionally.
(b) Strong scaling performance with fixed problem size while increasing the number of GPUs.}
\label{fig:scaling_results}
\end{figure}

To evaluate the scalability of the proposed framework, we conduct weak scaling experiments by increasing the time interval length $T$ alongside the number of GPUs, keeping the workload per GPU approximately constant. We show the weak scaling plot in Figure \ref{fig:scaling_results} (a) and the corresponding numerical results in Table~\ref{tab:weak_scaling}. As $T$ increases from $0.5$ to $4$ and the number of GPUs increases from $1$ to $8$, the total runtime grows modestly, increasing much more slowly than the problem size. This demonstrates solid weak scaling performance of the proposed time-parallel framework. We note that the number of iterations $l$ increases slightly from $3$ to $5$ as the problem size increases, which is expected since a longer time interval requires more iterations for the Parareal algorithm to converge (for a solution accuracy $\eta^k < 10^{-11}$). In all configurations, the pipelined scheduling consistently achieves shorter runtime than the regular scheduling. As the number of GPUs increases, synchronization overhead in the regular scheme becomes more significant, whereas the pipelined scheduling maintains higher resource utilization.
\begin{table}[H]
\caption{Weak scaling results}
\centering
\begin{tabular}{ccccc}
\hline
$T$ & GPU number $m$ & Regular time (s) & Pipeline time (s) & $l$ \\
\hline
0.5 & 1 & 4832.14 & 4283.23 & 3 \\
1   & 2 & 5122.68 & 4539.43 & 3 \\
2   & 4 & 7439.62 & 6416.24 & 4 \\
4   & 8 & 11745.27 & 9231.66 & 5 \\
\hline
\end{tabular}
\label{tab:weak_scaling}
\end{table}

\subsection{Strong Scaling}
To evaluate the strong scalability of the framework, strong scaling experiments are conducted by fixing the problem size and increasing the number of GPUs from 1 to 8. We show the strong scaling plot in Figure \ref{fig:scaling_results} (b) and the corresponding numerical results in Table~\ref{tab:strong_scaling}. The speedup and parallel efficiency are defined as
\begin{equation}
S_p = \frac{T_1}{T_p},\qquad E_p = \frac{S_p}{p} = \frac{T_1}{p\,T_p},
\label{eq:speedup_strong}
\end{equation}
where $T_1$ and $T_p$ denote the execution times using 1 and $p$ GPUs, respectively, and $p$ is the number of GPUs. When the number of GPUs increases from 1 to 4, the algorithm achieves near-linear speedup with parallel efficiency above 96\%. When scaling to 8 GPUs, the efficiency decreases to 77.1\%, mainly due to increased cross-node communication and synchronization overhead. Overall, the proposed time–space hybrid parallel framework demonstrates good scalability and computational efficiency on both single-node and multi-node GPU systems.

\begin{table}[htbp]
\centering
\caption{Multi-GPU strong scaling test}
\label{tab:strong_scaling}
\begin{tabular}{cccc}
\hline
GPU number & total time (s) & speedup & parallel efficient \\
\hline
1 & 9597.82 & - & 100\% \\
2 & 4834.20 & 1.99 & 99.5\% \\
4 & 2492.66 & 3.85 & 96.3\% \\
8 & 1555.91 & 6.17 & 77.1\% \\
\hline
\end{tabular}
\end{table}

\section{Discussion}
Building upon the experimental results presented above, we now discuss several key factors that determine the performance of the proposed framework, including the efficiency of spatial parallelism, the effectiveness of the pipelined temporal scheme, and the trade-offs associated with algorithmic parameters. 

The strong performance of GPU-based spatial parallelism is primarily due to the tailored mapping strategy between points and rods, along with the use of shared memory. In the current implementation, a system with 25 rods requires only about 2 GB of GPU memory, demonstrating good memory efficiency. Compared to the regular parallel-in-time method, the proposed pipeline approach significantly reduces GPU idle time. This advantage becomes more pronounced in longer simulations (i.e., larger $T$ in equation \eqref{eq:delta_w}), where waiting time accumulates in the regular scheme.

We find that the parameter $r$ plays a critical role in both performance and convergence. A larger $r$ leads to a faster coarse solver and can improve efficiency, but may degrade convergence as it yields solvers of lower accuracy. In particular, when rods are initially close to one another, the system becomes stiffer, and selecting a large $r$ may result in non-convergence. This highlights the need to balance computational efficiency with numerical stability in practice. In our framework, the number of time intervals $n$ (determined by the number of CPU cores) has limited impact on overall performance, as most of the computational workload is handled by the GPU. Additionally, the separation of computation and data transfer reduces communication overhead, contributing to strong scalability in practice.

Despite the advantages, the proposed framework has certain limitations. As the simulation time $T$ increases and more GPUs are employed, resource underutilization may occur, with some GPUs remaining idle during execution. This issue primarily stems from dependencies among time-parallel tasks and load imbalance. More efficient mapping and scheduling strategies between GPUs and solver tasks are therefore needed to further enhance performance. Additionally, the current Python-based implementation limits peak performance. Moving to lower-level languages could yield further speedups.

Overall, the experimental results show that a GPU-dominated framework leveraging both temporal and spatial parallelism can effectively minimize idle time while maximizing efficiency, suggesting a robust and scalable solution for space-time parallel simulations in heterogeneous computing environments.

\section{Conclusion}

In this work, we have presented a space–time parallel framework for simulating a fluid-structure interaction problem of filamentous swimmers on a heterogeneous CPU–-GPU architecture. The proposed method integrates spatial parallelism on the GPU with temporal parallelism based on the Parareal algorithm. A key contribution is the use of a pipeline structure to mitigate idle time in parallel-in-time execution, coupled with a GPU-friendly implementation of the underlying solver. Through a tailored mapping strategy and optimized memory usage, the proposed approach achieves high-intensity parallelization and efficient utilization of GPU resources.

Numerical results demonstrate that the pipeline method consistently outperforms the regular parallel-in-time scheme, particularly in long-time simulations. The method also exhibits favorable scalability and memory efficiency. These findings suggest that the synergistic combination of temporal and spatial parallelism offers an effective pathway to performance improvement for the long-time microswimmer dynamics simulations in biofluid research. Future work will focus on refining the GPU scheduling strategy and enhancing the robustness of the framework for stiff problems.

\section{Acknowledgements}
The 3D effect of the rod swimmer in the schematic illustration in the bottom panel of Figure 4 was enhanced using Gemini.

\bibliographystyle{IEEEtran}
\bibliography{references}

@article{Peskin2002,
  title = {The immersed boundary method},
  volume = {11},
  ISSN = {1474-0508},
  DOI = {10.1017/s0962492902000077},
  journal = {Acta Numerica},
  publisher = {Cambridge University Press (CUP)},
  author = {Peskin, Charles S.},
  year = {2002},
  month = jan,
  pages = {479–517}
}

@incollection{Olson2014,
  author = {Olson, S. D.},
  doi = {10.1090/conm/628/12544},
  editor = {Layton, A. T. and Olson, S. D.},
  booktitle = {Biological Fluid Dynamics: Modeling, Computations, and Applications},
  pages = {109--127},
  publisher = {AMS},
  series = {Contemporary Mathematics},
  title = {Motion of Filaments with Planar and Helical Bending Waves in a Viscous Fluid},
  volume = {628},
  year = {2014}
}

@article{tornberg2004simulating,
  author = {Tornberg, Anna-Karin and Shelley, Michael J.},
  title = {Simulating the dynamics and interactions of flexible fibers in Stokes flows},
  journal = {Journal of Computational Physics},
  volume = {196},
  number = {1},
  pages = {8--40},
  year = {2004},
  publisher = {Elsevier}
}

@article{Cortez2001,
  author = {Cortez, R.},
  title = {The Method of Regularized {S}tokeslets},
  journal = {SIAM. J. Sci. Comput.},
  year = {2001},
  volume = {23},
  pages = {1204--1225},
  number = {4},
  doi = {10.1137/S106482750038146X}
}

@article{carichino2019emergent,
  title={Emergent three-dimensional sperm motility: coupling calcium dynamics and preferred curvature in a Kirchhoff rod model},
  author={Carichino, Lucia and Olson, Sarah D},
  journal={Mathematical medicine and biology: a journal of the IMA},
  volume={36},
  number={4},
  pages={439--469},
  year={2019},
  publisher={Oxford University Press}
}

@article{ho2019three,
  title={A three-dimensional model of flagellar swimming in a Brinkman fluid},
  author={Ho, NguyenHo and Leiderman, Karin and Olson, Sarah},
  journal={Journal of Fluid Mechanics},
  volume={864},
  pages={1088--1124},
  year={2019},
  publisher={Cambridge University Press}
}

@article{lim2012fluid,
  title={Fluid-mechanical interaction of flexible bacterial flagella by the immersed boundary method},
  author={Lim, Sookkyung and Peskin, Charles S},
  journal={Physical Review E—Statistical, Nonlinear, and Soft Matter Physics},
  volume={85},
  number={3},
  pages={036307},
  year={2012},
  publisher={APS}
}

@ARTICLE{Lim2010,
  AUTHOR = {Lim, S.},
  TITLE = {Dynamics of an open elastic rod with intrinsic curvature and twist in a viscous fluid},
  JOURNAL = {Phys. Fluids},
  YEAR = {2010},
  volume = {22},
  number = {2},
  pages = {024104},
  doi = {10.1063/1.3326075}
}

@misc{lionsparareal,
  title={A Parareal in time discretization of PDEs. Comptes Rendus de l'Acad{\'e}, mie des Sciences--Series I--Mathematics 332 (7), 661--668 (2001)},
  author={Lions, JL and Maday, Y and Turinici, G}
}

@book{Pozrikidis1992,
  title = {Boundary Integral and Singularity Methods for Linearized Viscous Flow},
  ISBN = {9780511624124},
  DOI = {10.1017/cbo9780511624124},
  publisher = {Cambridge University Press},
  author = {Pozrikidis, C.},
  year = {1992},
  month = feb
}

@article{Rotne1969,
  author = {Rotne, J. and Prager, S.},
  title = {Variational treatment of hydrodynamic interaction in polymers},
  journal = {The Journal of Chemical Physics},
  volume = {50},
  number = {11},
  pages = {4831--4837},
  year = {1969},
  doi = {10.1063/1.1670977}
}

@article{Yamakawa1970,
  author = {Yamakawa, H.},
  title = {Transport properties of polymer chains in dilute solution: Hydrodynamic interaction},
  journal = {The Journal of Chemical Physics},
  volume = {53},
  number = {1},
  pages = {436--443},
  year = {1970},
  doi = {10.1063/1.1674100}
}

@article{Emmett2012,
  title = {Toward an efficient parallel in time method for partial differential equations},
  volume = {7},
  ISSN = {1559-3940},
  DOI = {10.2140/camcos.2012.7.105},
  number = {1},
  journal = {Communications in Applied Mathematics and Computational Science},
  publisher = {Mathematical Sciences Publishers},
  author = {Emmett, Matthew and Minion, Michael},
  year = {2012},
  month = mar,
  pages = {105–132}
}

@article{falgout2014parallel,
  title={Parallel time integration with multigrid},
  author={Falgout, Robert D and Friedhoff, Stephanie and Kolev, Tz V and MacLachlan, Scott P and Schroder, Jacob B},
  journal={SIAM Journal on Scientific Computing},
  volume={36},
  number={6},
  pages={C635--C661},
  year={2014},
  publisher={SIAM}
}

@article{gallagher2020passively,
  title={Passively parallel regularized stokeslets},
  author={Gallagher, Meurig T and Smith, David J},
  journal={Philosophical Transactions of the Royal Society A: Mathematical, Physical and Engineering Sciences},
  volume={378},
  number={2179},
  year={2020},
  publisher={The Royal Society}
}

@article{xue2024cpu,
  title={CPU--GPU heterogeneous code acceleration of a finite volume Computational Fluid Dynamics solver},
  author={Xue, Weicheng and Wang, Hongyu and Roy, Christopher J},
  journal={Future Generation Computer Systems},
  volume={158},
  pages={367--377},
  year={2024},
  publisher={Elsevier}
}

@article{liu2022parallel,
  title={Parallel-in-time simulation of biofluids},
  author={Liu, Weifan and Rostami, Minghao W},
  journal={Journal of Computational Physics},
  volume={464},
  pages={111366},
  year={2022},
  publisher={Elsevier}
}

@article{li2023incompressible,
  title={An incompressible flow solver on a GPU/CPU heterogeneous architecture parallel computing platform},
  author={Li, Qianqian and Li, Rong and Yang, Zixuan},
  journal={Theoretical and Applied Mechanics Letters},
  volume={13},
  number={5},
  pages={100474},
  year={2023},
  publisher={Elsevier}
}

@article{Ong2020PinTReview,
  author = {Ong, Benjamin W.},
  title = {A Review of Parallel-in-Time Algorithms},
  year = {2020}
}

@article{BlumersEtAl2021,
  author = {Blumers, Ansel L. and Yin, Minglang and Nakajima, Hiroyuki and Hasegawa, Yosuke and Li, Zhen and Karniadakis, George Em},
  title = {Multiscale parareal algorithm for long-time mesoscopic simulations of microvascular blood flow in zebrafish},
  journal = {Computational Mechanics},
  year = {2021},
  doi = {10.1007/s00466-021-02062-w}
}

@article{EGHBAL201757,
  title = {Acceleration of unsteady hydrodynamic simulations using the parareal algorithm},
  journal = {Journal of Computational Science},
  volume = {19},
  pages = {57-76},
  year = {2017},
  issn = {1877-7503},
  doi = {https://doi.org/10.1016/j.jocs.2016.12.006},
  author = {Araz Eghbal and Andrew G. Gerber and Eric Aubanel}
}

@article{Zeng2025,
  title = {A stable and efficient semi-implicit coupling method for fluid-structure interaction problems with immersed boundaries in a hybrid CPU-GPU framework},
  volume = {534},
  ISSN = {0021-9991},
  DOI = {10.1016/j.jcp.2025.114026},
  journal = {Journal of Computational Physics},
  publisher = {Elsevier BV},
  author = {Zeng, Yuhang and Wang, Yan and Yuan, Haizhuan},
  year = {2025},
  month = aug,
  pages = {114026}
}

@book{Higham2008,
  author = {Higham, Nicholas J.},
  title = {Functions of Matrices: Theory and Computation},
  publisher = {Society for Industrial and Applied Mathematics},
  address = {Philadelphia, PA},
  year = {2008},
  doi = {10.1137/1.9780898717778}
}

@article{olson2013modeling,
  title={Modeling the dynamics of an elastic rod with intrinsic curvature and twist using a regularized Stokes formulation},
  author={Olson, Sarah D and Lim, Sookkyung and Cortez, Ricardo},
  journal={Journal of Computational Physics},
  volume={238},
  pages={169--187},
  year={2013},
  publisher={Elsevier}
}

@article{Ho2001HyperactivationOM,
  title={Hyperactivation of mammalian spermatozoa: function and regulation.},
  author={Han-Chen Ho and Susan S. Suarez},
  journal={Reproduction},
  year={2001},
  volume={122 4},
  pages={519-26}
}

@article{Smith2009BendPI,
  title={Bend propagation in the flagella of migrating human sperm, and its modulation by viscosity.},
  author={David J. Smith and Eamonn A. Gaffney and Hermes Gad{\^e}lha and N. Kapur and Jackson C. Kirkman-Brown},
  journal={Cell motility and the cytoskeleton},
  year={2009},
  volume={66 4},
  pages={220-36}
}

@article{gander2026time,
  title={Time parallelization for hyperbolic and parabolic problems},
  author={Gander, Martin J. and Wu, Shu-Lin and Zhou, Tao},
  journal={Acta Numerica},
  year={2026},
  pages={1--},
  publisher={Cambridge University Press},
  note={arXiv preprint arXiv:2503.13526}
}

@article{hahne2021atmgrit,
  title={Asynchronous truncated multigrid-reduction-in-time (AT-MGRIT)},
  author={Hahne, Jens and Southworth, Ben and Friedhoff, Stephanie},
  journal={arXiv preprint arXiv:2107.09596},
  year={2021}
}

@article{freese2024psdc,
  title={Parallel performance of shared memory parallel spectral deferred corrections},
  author={Freese, Philip and G{\"o}tschel, Sebastian and Lunet, Thibaut and Ruprecht, Daniel and Schreiber, Martin},
  journal={arXiv preprint arXiv:2403.20135},
  year={2024},
  doi={10.48550/arXiv.2403.20135}
}

@article{steinstraesser2024pint,
  title={Parallel-in-time integration of the shallow water equations on the rotating sphere using Parareal and MGRIT},
  author={Steinstraesser, Jo{\~a}o Guilherme Caldas and Peixoto, Pedro da Silva and Schreiber, Martin},
  journal={Journal of Computational Physics},
  volume={496},
  pages={112591},
  year={2024},
  doi={10.1016/j.jcp.2023.112591}
}

@article{margenberg2021fsi,
  title={Parallel time-stepping for fluid–structure interaction},
  author={Margenberg, Nicolas and Richter, Thomas},
  journal={Computer Methods in Applied Mechanics and Engineering},
  volume={384},
  pages={113953},
  year={2021},
  doi={10.1016/j.cma.2021.113953}
}

\end{document}